\documentclass[superscriptaddress,aps,preprintnumbers,amsmath, amssymb, amsfonts,nofootinbib,10pt,twocolumn,prd, floatfix]{revtex4-2}
\usepackage{graphicx,subfigure, xcolor}
\usepackage[bookmarksopen,colorlinks=true,linkcolor=blue,citecolor=blue,urlcolor=blue]{hyperref}
\usepackage{comment}
\usepackage[inline]{enumitem}

\definecolor{purple}{rgb}{0.4 ,0, 0.85}

\usepackage{acro}
\DeclareAcronym{gr}{
    short=GR ,
    long=general relativity
}
\DeclareAcronym{ds}{
    short=dS ,
    long=de Sitter
}
\DeclareAcronym{ads}{
    short=AdS ,
    long=anti-de Sitter
}
\DeclareAcronym{qft}{
    short=QFT ,
    long={quantum field theory}
}
\DeclareAcronym{jt}{
    short=JT ,
    long={Jackiw-Teitelboim}
}

\DeclareAcronym{hl}{
    short=HL ,
    long=Ho\v{r}ava-Lifshitz
}

\DeclareAcronym{uv}{
    short=UV ,
    long=ultraviolet
}
\DeclareAcronym{ir}{
    short=IR ,
    long=infrared
}
\DeclareAcronym{wdw}{
    short=WDW ,
    long=Wheeler-DeWitt
}
\DeclareAcronym{rg}{
    short=RG ,
    long=renormalization group
}
\DeclareAcronym{adm}{
    short=ADM ,
    long={Arnowitt, Deser, and Misner}
}

\begin{document}
\preprint{RIKEN-iTHEMS-Report-24, STUPP-24-272, YITP-24-134}

\title{Jackiw-Teitelboim Gravity and Lorentzian Quantum Cosmology}

\author{Masazumi Honda}
\affiliation{Interdisciplinary Theoretical and Mathematical Sciences Program (iTHEMS), RIKEN, Wako 351-0198, Japan
}
\affiliation{Graduate School of Science and Engineering, Saitama University, 255 Shimo-Okubo,
Sakura-ku, Saitama 338-8570, Japan}

\author{Hiroki Matsui}
\affiliation{Center for Gravitational Physics and Quantum Information, 
Yukawa Institute for Theoretical Physics,
Kyoto University, 606-8502, Kyoto, Japan}

\author{Kota Numajiri}
\affiliation{Department of Physics, Nagoya University, Nagoya 464-8602, Japan}

\author{Kazumasa Okabayashi}
\affiliation{Center for Gravitational Physics and Quantum Information, 
Yukawa Institute for Theoretical Physics,
Kyoto University, 606-8502, Kyoto, Japan}

\begin{abstract}
We directly evaluate the probability amplitudes in Jackiw-Teitelboim (JT) gravity using the Lorentzian path integral formulation. By imposing boundary conditions on the scale factor and the dilaton field, the Lorentzian path integral uniquely yields the probability amplitude 
without contradiction. 
Under Dirichlet boundary conditions, we demonstrate that the amplitude derived from the Lorentzian path integral is expressed in terms of the modified Bessel function of the second kind. Furthermore, we provide the determinant for various boundary conditions and perform a detailed analysis of the Lefschetz thimble structure and saddle points.
In contrast to four-dimensional gravity, we show that the Hartle-Hawking no-boundary proposal is approximately valid in JT quantum cosmology. Furthermore, addressing quantum 
perturbation issues, we show that the quantum genesis of the two-dimensional universe occurs and exhibits perturbative regularity
when the dilaton field is non-zero and large as an initial condition.   
\end{abstract}

\date{\today}
\maketitle

\section{Introduction}

In recent years, there has been growing interest in two-dimensional gravity theories, particularly in \ac{jt}  gravity \cite{Jackiw:1984je, Teitelboim:1983ux}. The \ac{jt} gravity is a promising toy model for quantum gravity, as exactly solvable and renormalizable, unlike gravity theories in four or higher dimensions. Many studies focus on \ac{jt} gravity in \ac{ads} space \cite{Almheiri:2014cka, Jensen:2016pah, Maldacena:2016upp, Engelsoy:2016xyb}, which can be seen, e.g., as the near-horizon limit of an extreme black hole \cite{Almheiri:2014cka, Nayak:2018qej} (see also Refs.~\cite{Grumiller:2016dbn, Mertens:2022irh} for review).

Meanwhile, there has been progress in studying \ac{jt} gravity in \ac{ds} space, 
which is relevant to our universe due to the positive cosmological constant and curvature.
For \ac{ds} space, the canonical and path integral quantization of \ac{jt} gravity have been studied in the literature~\cite{Henneaux:1985nw, LouisMartinez:1993eh, Maldacena:2019cbz, Stanford:2019vob, Cotler:2019nbi, Iliesiu:2020zld, Stanford:2020qhm, Moitra:2021uiv, Anegawa:2023wrk, Nanda:2023wne,Buchmuller:2024ksd}. However, the equivalence between these two methods and the rigorous formulation of path integral in \ac{jt} gravity remains unclear.
In the path integral quantization of \ac{jt} gravity in \ac{ds} space, the probability amplitude analyses based on the Euclidean metric $g^{\rm E}_{\mu\nu}$, particularly following the Hartle-Hawking no-boundary proposal~\cite{Hartle:1983ai}, are commonly adopted.

The no-boundary proposal states that the wave function of the universe is given by the path integral over compact Euclidean geometries $g^{\rm E}_{\mu\nu}$, with the geometric configuration of spacetime that has a 3-dimensional configuration as the only boundary. This proposal elegantly explains the quantum birth of the universe from nothing~\cite{Vilenkin:1982de}. However, it has been criticized for various technical reasons, such as the convergence problems of the full integral over the Euclidean metric~\cite{Gibbons:1978ac}. In gravity, these path integrals correspond to excited states associated with negative energy eigenstates~\cite{Linde:1983mx}, leading to debates about the validity of the Euclidean gravity approach. 

To avoid these problems, the authors of Ref.~\cite{Halliwell:1988ik} proposed to perform path integrals along the steepest descent path of a complex metric, 
where the real part of the action decreases rapidly. 
In particular, recent studies in quantum cosmology are based on a rigorous method called the Lorentzian path integral formalism~\cite{Feldbrugge:2017kzv}. 
Although the integral of the phase factor $e^{\frac{i}{\hbar}S[g_{\mu\nu}]}$ is not manifestly convergent (conditionally convergent at most), deforming the integration contour to the complex plane using Picard-Lefschetz theory~\cite{Pham, Berry:1991, Howls, Witten:2010cx} enables us to rewrite it in terms of integrals along Lefschetz thimbles (steepest descents), which are absolutely convergent.\footnote{%
There are several studies on the quantum tunneling phenomenon, employing the Picard-Lefschetz theory~\cite{Mou:2019tck,Mou:2019gyl,Millington:2020vkg,Matsui:2021oio,Rajeev:2021zae,
Hayashi:2021kro,Feldbrugge:2022idb,Nishimura:2023dky,Feldbrugge:2023frq,Feldbrugge:2023mhn,Saito:2024acm}.} 
The Lorentzian path integral provides a consistent method for quantum gravity, allowing detailed insights into the wave function of the universe~\cite{Lehners:2023yrj,Lehners:2024kus}. 
Furthermore, utilizing the resurgence theory and Lefschetz thimble analyses in combination~\cite{Honda:2024aro},\footnote{A recent work~\cite{Chou:2024sgk} performed the Monte Carlo analysis using the generalized Lefschetz thimble method in Lorentzian quantum cosmology.} we can definitely perform the gravitational path integral over the Lorentzian spacetime and reduce the ambiguity of the wave function of the universe.

In this paper, we explore the quantum genesis of the universe within two-dimensional spacetime, focusing on \ac{jt} gravity as a renormalizable quantum gravity model~\cite{Jackiw:1984je,Teitelboim:1983ux} through the Lorentzian path integral. Quantum cosmology based on the Lorentzian path integral is expected to be studied within a consistent framework of quantum gravity. For instance, Lorentzian quantum cosmology has been discussed in \ac{hl} gravity~\cite{Matsui:2023hei}. However, implementing the Lorentzian path integral in gravity theories beyond \ac{gr} is usually challenging. In contrast, \ac{jt} gravity is renormalizable and allows for the straightforward implementation of the Lorentzian path integral. Therefore, although it is confined to two-dimensional spacetime, the quantum genesis of the universe can be analyzed within a consistent quantum gravity theory.%
\footnote{
The case of three-dimensional spacetime is discussed in \cite{chen_hikida_taki:2024short, chen_hikida_taki:2024long}.
}

The \ac{ds} space in \ac{jt} gravity is known to correspond to the Kantowski-Sachs model~\cite{Maldacena:2019cbz}, which represents a homogeneous four-dimensional universe with $S^1\times S^2$ topology. 
The Lorentzian path integral of the Kantowski-Sachs model has been analyzed in Refs.~\cite{Halliwell:1990tu,Fanaras:2021awm,Fanaras:2022twv,Ghosh:2023njl}.%
\footnote{The earlier works related to Kantowski-Sachs quantum cosmology were done in Refs.~\cite{Louko:1988bk,Conradi:1994yy,Louko:1988ia,Conti:2014uda}}
In particular, the authors of Refs.~\cite{Fanaras:2021awm, Fanaras:2022twv} 
performed this analysis under the Euclidean boundary condition following the original version of the no-boundary proposal and tunneling proposal. On the other hand, we directly evaluate the \ac{jt} probability amplitudes by performing the Lorentzian path integral, taking into account both the determinant and the physical boundary conditions where we mainly consider the Dirichlet boundary condition. Additionally, we provide a more detailed investigation of the thimble structure
and discuss the perturbation problems of Lorentzian quantum cosmology in \ac{jt} gravity. We show that the quantum genesis of the universe occurs and exhibits perturbative stability only when the dilaton field is non-zero and large as an initial condition.

The rest of the present paper is organized as follows. In Section~\ref{sec:WdW-wavefunctional}, we briefly review the canonical quantization of \ac{jt} gravity to compare the probability amplitude obtained from the Lorentzian path integral.
In Section~\ref{sec:Lorentzian-quantum-cosmology}, 
we analyze the probability amplitude of \ac{jt} gravity under the Dirichlet boundary condition utilizing the Lorentzian path integral approach. We show the amplitude derived from the path integral coincides with the minisuperspace solutions of the canonical quantization.
We also provide the Lefschetz thimble structure and saddle point analysis in detail. The case with the Neumann boundary condition is discussed in Section~\ref{sec:Neumann-boundary-condition}.
In Section~\ref{sec:perturbation-problems}, we consider the problem of perturbations in quantum cosmology in the \ac{jt} gravity and show that 
the perturbative regularity is preserved for the quantum genesis of the universe 
in the \ac{jt} gravity. 
Section~\ref{sec:conclusions} is devoted to conclusions. The Appendix~\ref{sec:appendix-prefactor} exhibits detailed calculations of the amplitude with precise prefactors under various boundary conditions.

\section{Canonical quantization of Jackiw-Teitelboim gravity}
\label{sec:WdW-wavefunctional}

In this section, we review the canonical quantization of \ac{jt} gravity, 
following the method outlined by Refs.~\cite{Henneaux:1985nw, LouisMartinez:1993eh,Iliesiu:2020zld}, and derive the probability amplitude of the universe to compare it with the results obtained from the Lorentzian path integral approach in the next section. 
Consider the action of \ac{jt} gravity in Lorentzian signature,
\begin{equation}
\label{eq:action-jt-general}
S_{\text{JT}} = \frac{1}{2}\int_{\mathcal{M}}\mathrm{d}^2 x \sqrt{-g}
\phi[R - 2\Lambda] - \int_{\partial \mathcal{M}} \mathrm{d}y \sqrt{\gamma}\phi\,\mathcal{K},
\end{equation}
where the gravitational constant is normalized ($\frac{\phi}{8\pi G}\to \phi$). The dilaton field should be non-negative, $\phi \ge 0$, since a negative value results in an effective Newton constant $\frac{\phi}{8\pi G}$ that is also negative. This bound is also reasonable when considering the JT theory arising from the reduction of 4-dimensional spacetime.
Here, $\Lambda$ denotes a positive cosmological constant, $g$ represents the two-dimensional spacetime metric on $\mathcal{M}$, and $\sqrt{\gamma}$ is the determinant of the induced metric $\gamma_{ab}$ on the boundary $\partial \mathcal{M}$. The boundary term in the action is crucial for the variational principle to be applicable under Dirichlet boundary conditions for both the metric and dilaton fields.

We will use the following \ac{adm} decomposition of the metric 
\begin{align}
\mathrm{d}s^2
=-N^2(x,t) \mathrm{d}t^2+g_{1}(x,t)
(\mathrm{d}x+N_{\perp}(x,t)\mathrm{d}t)^2\,,\label{adm}
\end{align}
where $N$ is the lapse, $N_{\perp}$ is the shift, 
and $g_{1}$ is the spatial metric. Then, $t$ represents Lorenzian time, and $x$ indicates the spatial direction within the range $0\leq x<2\pi$. 
After integrating by parts and using the boundary terms with the extrinsic curvature scalar $\mathcal{K}=\frac{1}{2Ng_{1}}\left(\dot{g_{1}}-2N_{1|1}\right)=\frac{1}{2Ng_{1}}\left(\dot{g_{1}}-2N_{\perp}^\prime g_{1}-N_{\perp} g_{1}^{\prime}\right)$ where a vertical bar denotes covariant differentiation with respect to the spatial metric and $N_1=g_1N_{\perp}$, 
the action can then be written as,
\begin{align}
&S_{\text{JT}}=\int \mathrm{d}^2x \left[\frac{\dot{\phi}}{N}\left(\frac{N_{\perp}}{2\sqrt{g_1}}{g_1'}+\sqrt{g_1} N_{\perp}'-\frac{\dot{g}_1}{2\sqrt{g_1}}\right)\right]\nonumber\\
&+\int \mathrm{d}^2x\Biggl[ \frac{\phi'}{N}\left(\frac{NN'}{\sqrt{g_1}}-\sqrt{g_1}N_{\perp}N_{\perp}'+\frac{N_{\perp}}{2\sqrt{g_1}}\dot{g}_1-\frac{N_{\perp}^2}{2\sqrt{g_1}}g_1'\right)\nonumber \\
&\quad -N\sqrt{g_1}\phi\Lambda\Biggr]\,,
\label{jt-action}
\end{align}
where the dots denote derivatives with respect to $t$ and the primes denote derivatives with respect to $x$. As usual, the action does not involve time derivatives of fields $N$ and $N_{\perp}$ and therefore 
\begin{equation}
\Pi_{N} = \Pi_{N_{\perp}} =0,
\end{equation}
which act as primary constraints. 
The momenta conjugate to the dilaton and scale factor are,
\begin{align}
\Pi_{g_1} &=-\frac{\dot{\phi}}{2N\sqrt{g_1}}+\frac{N_{\perp}\phi'}{2N\sqrt{g_1}},\\
\Pi_\phi &= \frac{N_{\perp}}{2N\sqrt{g_1}}g_1'+\frac{N_{\perp}'\sqrt{g_1}}{N}-\frac{\dot{g}_1}{2N\sqrt{g_1}}.
\end{align}

The canonical Hamiltonian is,
\begin{equation}
H = \int \mathrm{d}x \left[ N \mathcal{H} + N_{\perp} \mathcal{P} \right]\,,
\end{equation}
where 
\begin{align}
\mathcal{H}
&=2\Pi_\phi \Pi_{g_1}\sqrt{g_1}-\left(\frac{\phi'}{\sqrt{g_1}}\right)'-\sqrt{g_1}\phi\Lambda \,,
\label{hamiltonian-constraint}\\
\mathcal{P}
&=2g_1\Pi_{g_1}'+\Pi_{g_1}g_1'-\Pi_\phi \phi'\,,
\label{momentum-constraint}
\end{align}
and the Hamiltonian and momentum constraints are respectively $\mathcal{H}\approx 0$
and $\mathcal{P} \approx 0$ in the Dirac sense.
So far, our discussion has focused on classical approaches. We now move to the quantum theory, introducing fields as operators. In this framework, canonical quantization is performed formally by replacing the canonical momenta as
\begin{equation}
\hat{\Pi}_{g_1} = - i\hbar \frac{\delta}{\delta g_1(x)},~~~
\hat{\Pi}_\phi = - i\hbar \frac{\delta}{\delta \phi(x)}\,,
\end{equation}
and the state is represented by wave functional denoted as $\Psi[g_1, \phi]$. 
In the quantum theory of gravity, 
the physical wave function $\Psi[g_1, \phi]$ has to satisfy 
the Hamiltonian and momentum constraints respectively, 
\begin{align}\label{wheeler-dewitt}
\mathcal{H}\, \Psi[g_1, \phi]=0,~~~\mathcal{P}\, \Psi[g_1, \phi]=0\,,
\end{align}
where the Hamiltonian constraint equation is also referred to as the \ac{wdw} equation.

\subsection{General Solutions}

Following Refs.~\cite{Henneaux:1985nw, LouisMartinez:1993eh},
we can eliminate the momenta conjugate to the dilaton $\Pi_\phi$
by using a linear combination of the Hamiltonian
and Momentum constraints, 
\begin{align}
0 &\approx 2 \Pi_{g_1} \sqrt{g_1} \mathcal{P}+\phi^{\prime} \mathcal{H} \notag \\
& =2 \sqrt{g_1}\left(g_1 \Pi_{g_1}^2-\frac{1}{4}\left(\frac{\phi^{\prime}}{\sqrt{g_1}}\right)^2-\frac{\Lambda}{4} \phi^2\right)^{\prime}\,,
\end{align}
and we have 
\begin{equation}\label{hamiltonian-momentum-constraint}
4g_1 \Pi_{g_1}^2=\left(\frac{\phi^{\prime}}{\sqrt{g_1}}\right)^2+\Lambda\phi^2 -C\,,
\end{equation}
where $C$ is the constant for the on-shell configurations. 
Now, it is found that 
solving constraints~\eqref{wheeler-dewitt} is equivalent to imposing the above conditions~\eqref{hamiltonian-momentum-constraint} 
and the momentum constraint~\eqref{momentum-constraint} in the Dirac sense.
In the quantum theory, we shall impose 
\begin{equation}
\hat{\Pi}_{g_1}\Psi[g_1, \phi]= \pm\frac{\sqrt{\left(\frac{\phi^{\prime}}{\sqrt{g_1}}\right)^2+\Lambda\phi^2 -C}}{2\sqrt{g_1}}
\Psi[g_1, \phi]\,.
\end{equation}
Thus, we have the following solution~\cite{Henneaux:1985nw, LouisMartinez:1993eh}
\begin{align}\label{eq:wheeler-dewitt-solution}
&\Psi_{\pm}\left[g_1,\phi\right]=e^{\pm \frac{i}{\hbar}\int d x\left(Q-\phi^{\prime}(x) \tanh ^{-1}\left(\frac{Q}{\phi^{\prime}(x)}\right)\right)}\,, \\
&\quad Q  =\sqrt{\left(\Lambda\phi(x)^2-C\right) g_1(x)+\phi^{\prime}(x)^2}\,,
\end{align}
where it can be shown that the above wave function satisfies the momentum constraint $\mathcal{P}\Psi_{\pm}[g_1, \phi]=0$~\cite{Nanda:2023wne}.
The general solutions consist of a linear combination of 
$\Psi_{+}\left[g_1,\phi\right]$ and $\Psi_{-}\left[g_1,\phi\right]$, which describe the contracting and expanding configurations, respectively.

More general solutions would be given as a sum over various constant $C$ and therefore
can be written as~\cite{Iliesiu:2020zld},
\begin{align}\label{eq:general-solution}
&\Psi\left[g_1,\phi\right]=\int \mathrm{d}C\tilde{\rho}(C)
e^{+ \frac{i}{\hbar}\int \mathrm{d} x\left(Q-\phi^{\prime}(x) \tanh ^{-1}\left(\frac{Q}{\phi^{\prime}(x)}\right)\right)} \notag \\
&\quad +\int \mathrm{d}C\rho(C)
e^{-\frac{i}{\hbar}\int \mathrm{d} x\left(Q-\phi^{\prime}(x) \tanh ^{-1}\left(\frac{Q}{\phi^{\prime}(x)}\right)\right)}\,,
\end{align}
where $\tilde{\rho}(C),\rho(C)$ are arbitrary complex functions of $C$.
The norm of the wave function is not necessarily positive or conserved for all solutions of the \ac{wdw} equation. Therefore, it has been argued that $\tilde{\rho}(C)$ and $\rho(C)$ should be chosen in a way that ensures both a positive and conserved norm (see, for instance, Ref.~\cite{Maldacena:2019cbz,Nanda:2023wne} for a detailed discussion).

The solution~\eqref{eq:general-solution} involves arbitrariness, making it difficult to determine which solution represents a physical one, especially in the absence of well-defined boundary conditions for the \ac{wdw} equation. Ref.~\cite{Maldacena:2019cbz} suggests that the wave function of \ac{ds} spacetime should consist of $\Psi_{-}\left[g_1,\phi\right]$: this describes the expanding configurations in the large $g_1,\phi$ limit. Thus, we can only take $\Psi_{-}\left[g_1,\phi\right]$, and $\rho(C)$ is chosen from some prescriptions. According to Ref.~\cite{Iliesiu:2020zld}, $\rho(C)$ 
is determined as $\rho(C)=\sinh (2\pi \sqrt{C})$ by requiring that the wave function in this limit corresponds to the Schwarzian partition function.
Consequently, when we assume the dilaton profile $\phi$ along a spatial slice with $\mathrm{d}u=\sqrt{g_1}\mathrm{d}x$, and denote $l=\int \sqrt{g_1}d x$, 
the general solution results in the following form,
\begin{align}\label{eq:wdw-general-solution1}
&\Psi[l,\phi]=\int^{\infty}_{0} \mathrm{d}C\sinh (2\pi \sqrt{C}) \notag\\
&\times e^{-\frac{i}{\hbar}\int^{l}_{0} \mathrm{d}u\left(
\sqrt{\Lambda\phi^2-C+(\partial_u\phi)^2}-\partial_u\phi \tanh ^{-1}\left(\sqrt{1+\frac{\Lambda\phi^2-C}{(\partial_u\phi)^2}}\right)\right)} \notag \\
&=\int^{\infty}_{0} \mathrm{d}C\sinh (2\pi \sqrt{C})e^{-\frac{il}{\hbar}\sqrt{\Lambda\phi^2-C}}\,,   
\end{align}
where we take $\partial_u\phi=0$ in the last expression. 
Then, by performing the integral we obtain~\cite{Iliesiu:2020zld}, 
\begin{equation}\label{eq:wdw-general-solution2}
\Psi[l,\phi]=\frac{l\Lambda\phi^2/\hbar}{l^2/\hbar^2-4\pi^2} K_{2}\left(i\sqrt{\Lambda\phi^2(l^2/\hbar^2-4\pi^2)}\right)\,,  
\end{equation}
where $K_{\nu}(z)$ is the modified Bessel function of the second kind. 
In large $l,\phi$ limit, we obtain
\begin{equation}\label{eq:wdw-minisuperspace-solution1}
\Psi\left[l,\phi\right] \propto e^{-\frac{il}{\hbar}\sqrt{\Lambda\phi^2}}\,.
\end{equation}
Ref.~\cite{Iliesiu:2020zld} also proposes another wave function written in terms of the modified Bessel function of the first kind $I_{\nu}(z)$ to realize the real wave function following the original Hartle-Hawking proposal.
From the above discussion, the \ac{jt} wave function is expected to be given by specific Bessel functions. 
In the next subsection, we will confirm that these results remain valid even if we take the minisuperspace approximation from the first place.

\subsection{Minisuperspace Canonical Quantization}

Hereafter, we shall evaluate the above results~\eqref{eq:wdw-general-solution2} or \eqref{eq:wdw-minisuperspace-solution1} by starting from the minisuperspace approximation. In the minisuperspace, 
$g_1$, $\phi$ are independent of $x$, and we take the following ADM metric,
\begin{equation}\label{eq:flrw}
\mathrm{d}s^2 =- N(t)^2 \mathrm{d}t^2 + a(t)^2 \mathrm{d}x^2\,.
\end{equation}
In this setup, the action for \ac{jt} gravity reduces 
\begin{align}
S_{\rm JT}
&=-2\pi\int\mathrm{d}t \left(\frac{\dot{q}\dot{\varphi}}{4N}
+N\Lambda \right)\,,\label{eq:action-jt-minisuperspace}
\end{align}
where we transformed the lapse function $N\to N/a\phi$, 
\footnote{Shifting the lapse function corresponds to introducing a new time coordinate, $\tau$ which is related by $\mathrm{d}t=\frac{\mathrm{d}\tau}{a\phi}$. 
}
and defined $q(t)=a^2(t)$ and $\varphi(t)=\phi^2(t)$ for simplicity.
Performing the canonical quantization, the canonical conjugate momenta are transformed into Hermitian operators, 
\begin{align}
\begin{split}
&\Pi_{q} = - \pi\frac{\dot{\varphi}}{2N},~~~
\Pi_{\varphi} = - \pi\frac{\dot{q}}{2N}, \\
& \Longrightarrow \quad 
\hat{\Pi}_{q} = - i\hbar\frac{\partial}{\partial q},~~~
\hat{\Pi}_{\varphi} = - i\hbar\frac{\partial}{\partial \varphi}\,.    
\end{split}
\end{align}
The minisuperspace \ac{wdw} equation for the \ac{jt} gravity is 
written as 
\begin{align}\label{eq:wheeler-dewitt-minisuperspace}
\left[\frac{\partial}{\partial \tilde{q}}\frac{\partial}{\partial \tilde{\varphi}} + 1\right]\Psi[\tilde{q}, \tilde{\varphi}]=0\,,
\end{align}
where $\tilde{q}=\frac{\pi q}{\hbar}\sqrt{\Lambda}$ and 
$\tilde{\varphi}=\frac{\pi \varphi}{\hbar}\sqrt{\Lambda}$.
The general solution of the minisuperspace \ac{wdw} equation~\eqref{eq:wheeler-dewitt-minisuperspace} is given by a linear combination of a function $\left({\tilde{q}}/{\tilde{\varphi}}\right)^{\frac{m}{2}}F_m(2\sqrt{\tilde{q}\tilde{\varphi}})$~\cite{Maldacena:2019cbz} where $F_m(z)$ is a specific Bessel function and $m$ is the arbitrary real number. Ref.~\cite{Fanaras:2021awm} showed that under the out-going mode assumption describing expanding universe in large $a,\phi$ limit,
the appropriate Bessel functions are the Hankel functions of the second kind $H_m^{(2)}(z)$.
Thus, the wave function should take the following form,
\begin{equation}\label{eq:wdw-minisuperspace-solution2}
\Psi_m[q, \varphi]=
\left(\frac{q}{\varphi}\right)^{\frac{m}{2}} H_m^{(2)} \left(\frac{2\pi}{\hbar}\sqrt{\Lambda q\varphi}\right).
\end{equation}
which asymptotes
\begin{equation}\label{eq:wdw-minisuperspace-solution3}
\Psi_m\left[q\varphi\gg \frac{1}{\Lambda}\right]
\propto e^{-
\frac{2\pi i}{\hbar}\sqrt{\Lambda q\varphi}}\,,
\end{equation}
reproducing Eq.~\eqref{eq:wdw-minisuperspace-solution1}. Assuming correspondence with the Schwarzian theory~\cite{Stanford:2017thb} in large $a,\phi$ limit forces $m$ to be $-2$.
On the other hand, Ref.~\cite{Iliesiu:2020zld} suggests that $F_m(z)$ should be taken as either $K_m(z)$ or $I_m(z)$ as previously discussed. Actually, it is impossible to determine which Bessel function is appropriate, and the value of $m$, solely by solving the \ac{wdw} equation~\eqref{eq:wheeler-dewitt-minisuperspace}. In contrast, when $(q_0, \varphi_0)$ and $(q_1, \varphi_1)$ are fixed, the path integral approach provides a more straightforward way to obtain the physical probability amplitude. In the following section, we directly evaluate the \ac{jt} probability amplitudes by performing the Lorentzian path integral and demonstrate the exact expression for the wave function written in terms of the modified Bessel function of the second kind, $K_0(z)$.

\section{Lorentzian quantum cosmology}
\label{sec:Lorentzian-quantum-cosmology}

Hereafter, we perform the path integral quantization of \ac{jt} gravity and
analyze the probability amplitude of the universe utilizing the Lorentzian path integral approach. 
The probability amplitude of \ac{jt} gravity can be schematically 
given by the gravitational path integral
\begin{equation}
G[g,\phi]= \int_{{\cal M}}\mathcal{D}g_{\mu\nu}\mathcal{D}\phi\, \exp \left(\frac{i}{\hbar}S_{\text{JT}}[g_{\mu\nu},\phi]\right) \,.
\end{equation}
In the minisuperspace approximation, 
utilizing the \ac{jt} action~\eqref{eq:action-jt-minisuperspace}, we can derive the Lorentzian probability amplitude as~\cite{Halliwell:1988wc}%
\footnote{Here $q, \varphi$ run from $-\infty$ to $\infty$ although they are originally positive. This treatment can avoid configurational difficulties from undetermined classical paths, which can happen when they are limited to positive domains \cite{Halliwell:1988wc}. 
This extension does not alter physical implications as long as we choose boundary conditions consistent with their positivities.}
\begin{align}\label{Lorentzian-amplitude}
G\left[q,\varphi\right] = 
\int\! \mathrm{d}N(t_f-t_i) \int \mathcal{D}q\mathcal{D}\varphi 
\exp\left(iS_{\rm JT}[N,q,\varphi]/\hbar\right)\,,
\end{align}
where we proceed with the lapse integral and the path integral over all configurations of the scale factor $q(t)$ and dilaton field $\varphi(t)$ 
with the given initial $(t_i=0)$ and final $(t_f=1)$ values. 
In this paper, we mainly consider the integration of the lapse function over $N \in (0, \infty)$, and this choice of $N$ ensures the causality~\cite{Teitelboim:1983fh}. Although one can also consider the integration on the entire real axis, $N \in (-\infty, \infty)$, this leads to the ambiguity of the wave function~\cite{DiazDorronsoro:2017hti,Feldbrugge:2017mbc,Honda:2024aro}.

First, we assume the Dirichlet boundary condition in the Lorentzian path integral, where we fix the value of the scale factor and dilaton field at the two endpoints,
\footnote{When adopting the no-boundary prescription, the field value of the dilaton becomes imaginary in the Euclidean region~\cite{Maldacena:2019cbz}. Consequently, to precisely reproduce the no-boundary wave function via the Lorentzian path integral, it is necessary to set the initial condition of the dilaton field $\varphi$ to be imaginary.}  
\begin{align}\label{eq:Dirichlet-boundary-condition}
\begin{split}
q(t_{i}=0)&=q_{0}, \ \varphi(t_{i}=0)=\varphi_{0}, \\
q(t_{f}=1)&=q_{1}, \ 
\varphi(t_{f}=1)=\varphi_{1}.   
\end{split}
\end{align} 
The equations of motion obtained by varying the \ac{jt} action~\eqref{eq:action-jt-minisuperspace} 
with respect to $q(t)$ and $\varphi(t)$ are
\begin{equation}
\ddot q=0, \quad
\ddot \varphi=0\,,
\end{equation}
and we can get the analytical solutions of $q(t)$ and $\varphi(t)$ with the Dirichlet boundary condition~\eqref{eq:Dirichlet-boundary-condition}.

For the \ac{jt} action~\eqref{eq:action-jt-minisuperspace}, the path integral  under the Dirichlet boundary condition can be exactly evaluated, and we find the following expression (see Appendix~\ref{sec:appendix-prefactor}. for the detail derivation),
\begin{align}\label{G-amplitude-Dirichlet}
G\left[q_{1},\varphi_{1};q_{0},\varphi_{0}\right]
= \int_{0}^\infty \! \frac{\mathrm{d}N }{4\hbar N}
\exp \left(\frac{i S_\textrm{on-shell}[ N]}{\hbar}\right),
\end{align}
where the on-shell action
$S_\textrm{on-shell}[N]$ is written as,%
\footnote{The on-shell action
$S_\textrm{on-shell}[N]$ exactly corresponds with the on-shell action of the Kantowski-Sachs model, and the steepest-descent analysis has been conducted in Ref.~\cite{Halliwell:1990tu}. However, the authors assume an Euclidean metric and do not specify the Lorentzian integration contour for the lapse function.}
\begin{equation}\label{eq:on-shell-action}
S_\textrm{on-shell}[N] 
=-\beta N - \frac{\gamma}{N} \,, 
\end{equation}
with 
\begin{equation}
\beta =2\pi \Lambda, \quad \gamma = 
\frac{\pi}{2}(q_1 -q_0 )(\varphi_1 -\varphi_0 )\,.   
\end{equation}

Now, we consider the integral by using the steepest descents from saddle points.
For this purpose, it is convenient to work in the following expression,
\begin{equation}
G(\hbar) 
= \frac{1}{4\hbar}\int_{-\infty}^\infty  \! \mathrm{d}x \exp [F_D(x)] \,,
\label{eq:integralFD}
\end{equation}
where we took $N=e^x$ and introduced 
\begin{equation}\label{eq:exponential-part1}
F_D(x) := \frac{i}{\hbar} S_\textrm{on-shell}(x)
=\frac{i}{\hbar}\left[-\beta e^x - \frac{\gamma}{e^x}\right]\,,
\end{equation}
as the exponential part of the above integration.
The saddle points are obtained by requiring that the derivative of the exponent in the action vanishes, i.e., $\frac{\mathrm{d}F_D(x)}{\mathrm{d}x}=0$: 
\begin{align}
    x_{s \pm}^{(m)} = \log \left(\pm \sqrt{\frac{\gamma}{\beta}}\right)+2m \pi i  
    \ \Longrightarrow \ N_{s \pm}=\pm \sqrt{\frac{\gamma}{\beta}}\,,
\label{eq:saddles}
\end{align}
where $m \in \mathbb{Z}$, and we take the branch cut of the logarithmic function as the negative real axis such that $\log(-1)= -\pi i$.

It is a priori nontrivial whether or not complex saddle points contribute to the integral. A systematic way to see this is to find Lefschetz thimbles (steepest descents) associated with the saddles and then appropriately perform deformation of the integral contour to a superposition of the thimbles equivalent to the original integral.
Important properties of the Lefschetz thimble $\mathcal{J}_{x_s}$ associated with the saddle point $x_s$ are (i) $\textrm{Im}[F_D(x)] =\textrm{Im}[F_D(x_s)]$ along $x \in \mathcal{J}_{x_s}$ and (ii) monotonically decreasing $\textrm{Re}[F_D(x)]$ as we get away from $x_s$ along $x \in \mathcal{J}_{x_s}$.
Then one can rewrite the integral as
\begin{equation}
G(\hbar)
=\frac{1}{4\hbar} \sum_{x_{s}} n_{x_s} \int_{\mathcal{J}_{x_s}} \mathrm{d} x \exp [F_D (x)] \,, 
\end{equation}
where $n_{x_s}$ called Stokes multiplier is an integer to determine how the saddle $x_s$ contributes to the integral.
While the integer $n_{x_s}$ cannot change continuously as we change the parameters,
it may discontinuously jump across particular regions of the parameters called Stokes lines.
In this situation, the $\hbar$ expansion gets a change of its form as varying the parameters.
This is called the Stokes phenomenon that can occur when there are multiple saddle points with $\textrm{Im}[F_D (x_s)]=\textrm{Im}[F_D (x_s')]$ for $x_s \neq x_s'$.
It is known that on the Stokes lines, the Lefschetz thimbles pass multiple saddle points and the Stokes multiplier $n_{x_s}$ is not unique.
Therefore, the decomposition by the thimbles becomes ambiguous on the Stokes lines.
In this situation, it is often useful to deform the parameters slightly away from the Stokes lines and try to take limits to approach the Stokes lines from different directions.
There are various known examples where the thimble decomposition has terms with jumps of the Stokes multiplier across the Stokes lines while the total answer is continuous across the Stokes lines because of cancellation of the ambiguities against other ambiguities coming from the subtlety of resummation of perturbative series \cite{Honda:2024aro}. 

%
%
\begin{figure}[t] 
\includegraphics[width=0.48\textwidth]{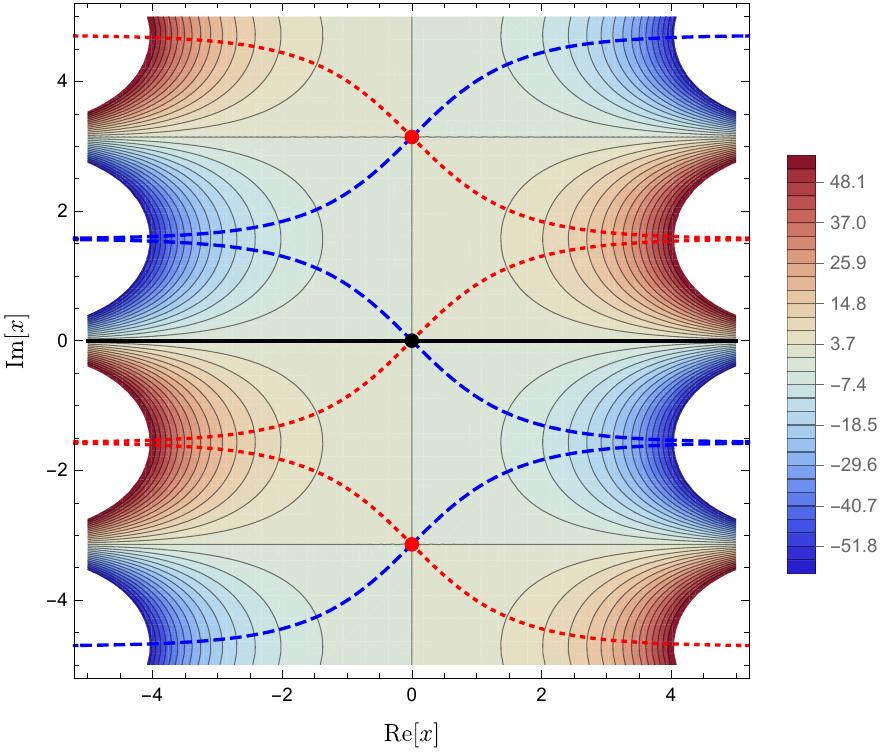}
\caption{The contour plot of $\textrm{Re}\left[F_D(x)\right]$ over the complex $x$ plane for $\beta=1$, $\gamma=1$, and $\hbar =1$ as a representative of the case with the same signs of $\beta$ and $\gamma$. 
The black (or red) circles correspond to $x^{(m)}_{s+}$ (or $x^{(m)}_{s-}$), while the blue dashed (or red dotted) lines represent the Lefschetz thimbles (or their duals). The black horizontal line indicates the original integration contour.}
\label{fig:Picard-Lefschetz1}
\end{figure} 
%

In our case, we have the two parameters and can consider two possible cases; the parameters $\beta$ and $\gamma$ have the same or opposite signs. 
In the case where the parameters have the same signs, the steepest-descent contours are shown in Fig.~\ref{fig:Picard-Lefschetz1}, and we can find that the contributing saddle point is given by $x^{(0)}_{s+}= \frac{1}{2} \log \left| \frac{\gamma}{\beta} \right|$ without any ambiguities.
Its associated thimble is equivalent to the original integral contour $\mathbb{R}$ by the Cauchy integration theorem.
Then, changing the integral variable as $x=x_s+\delta x$, we can perform the path integral,
\begin{align}
    G(\hbar) 
    &= \frac{1}{4\hbar}\int_{-\infty}^\infty  \! \mathrm{d}\delta x \exp [F_D(x^{(0)}_{s+}+\delta x)] \notag \\
    &=\frac{1}{4\hbar}\int_{-\infty}^\infty  \! \mathrm{d}\delta x \exp \left[-\frac{2i}{\hbar}\sqrt{\beta\gamma } \cosh ({\delta x})\right]\notag \\
    & = \frac{1}{2\hbar }K_0\left(\frac{2i\sqrt{\beta\gamma }}{\hbar }\right) 
    = - \frac{\pi i}{4\hbar }H^{(2)}_0\left(\frac{2\sqrt{\beta\gamma }}{\hbar }\right)\,,
\label{eq:G_K0}
\end{align}
where $K_{\nu}(z)$ has the following integral representation for $\textrm{Re}[z]>0$,\footnote{
Our case corresponds to $\textrm{Re}[z]=0$ and hence \eqref{eq:Bessel} is not directly available while \eqref{eq:G_K0} is correct as a result.
To see this, one first rewrites the second line of \eqref{eq:G_K0} as $G(\hbar ) =-\frac{\pi}{2}\left( Y_0 (2\sqrt{\beta\gamma}/\hbar ) +i J_0 (2\sqrt{\beta\gamma}/\hbar ) \right)$, where the Bessel functions $Y_0 (z)$ and $J_0 (z)$ have the integral representations: $Y_0 (z)=\int_0^\infty ds \cos{(z\cosh{s})}$ and $J_0 (z)=\int_0^\infty ds \sin{(z\cosh{s})}$,
which are convergent for real $z$.
Then \eqref{eq:G_K0} is derived by using the formula 
$K_\nu (iz) =-\frac{\pi}{2i^\nu}\left( Y_\nu (z) +i J_\nu (z)  \right) = -\frac{\pi}{2i^{\nu-1}} H^{(2)}_{\nu}(z)$.}
\begin{equation}
K_{\nu}(z)=\frac{1}{2}\int_{-\infty}^\infty \! \mathrm{d}s 
e^{-\nu s- z\cosh ({s})} \,. 
\label{eq:Bessel}
\end{equation}
Thus, the probability amplitude~\eqref{eq:integralFD} satisfies the minisuperspace \ac{wdw} equation~\eqref{eq:wheeler-dewitt-minisuperspace}.
The asymptotic form of $G(\hbar) $ with $\hbar \to 0$ leads to
\begin{equation}
G(\hbar) \simeq 
\frac{\pi^{\frac{1}{2}}}{4\hbar^{\frac{1}{2}}(\beta\gamma)^{\frac{1}{4}} }
e^{-\frac{2i}{\hbar}\sqrt{\beta\gamma } -\frac{\pi i}{4}
}\,, 
\end{equation}
which derives $G\left[q_{1},\varphi_{1}\right]\propto e^{-
\frac{2\pi i}{\hbar}\sqrt{\Lambda q_1\varphi_1}}$ to be consistent with 
the solution~\eqref{eq:wdw-minisuperspace-solution1}.
The existence of the real saddle point implies that the path integral describes classical evolution from $(q_0,\varphi_0)$ to $(q_1,\varphi_1)$. In particular, if we consider the scenario where $q_0 = \varphi_0 = 0$, the classical initial singularity cannot be avoided, and perturbative effects cause the probability amplitude to become ill-defined, as we will demonstrate in Section~\ref{sec:perturbation-problems}.

%
\begin{figure}[t] 
\includegraphics[width=0.48\textwidth]{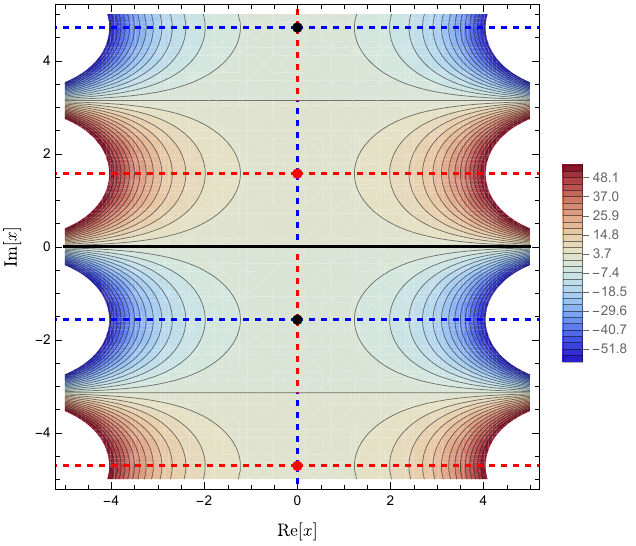}
\caption{A similar plot to Figs.~\ref{fig:Picard-Lefschetz1} with $\beta=1$ and $\gamma=-1$, and $\hbar =1$ as a representative of the case with the opposite signs of $\beta$ and $\gamma$}.
\label{fig:Picard-Lefschetz2}
\end{figure} 
%
%
\begin{figure}[t] 
\subfigure[$\theta =+\frac{\pi }{10}$]{%
\includegraphics[width=0.48\textwidth]{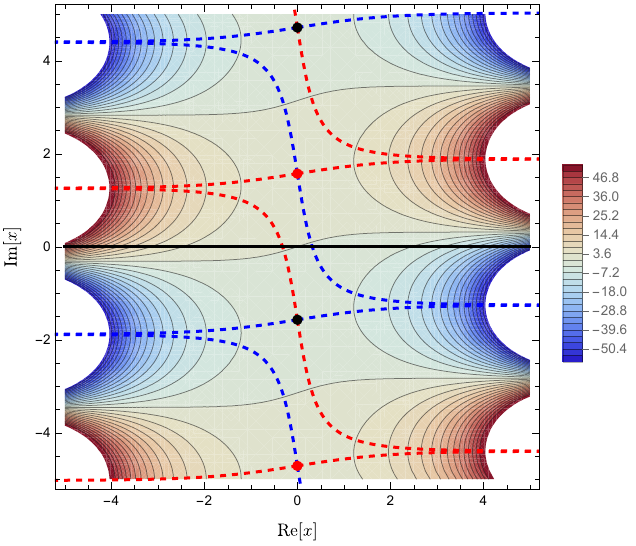}}
\subfigure[$\theta =-\frac{\pi }{10}$]{%
\includegraphics[width=0.48\textwidth]{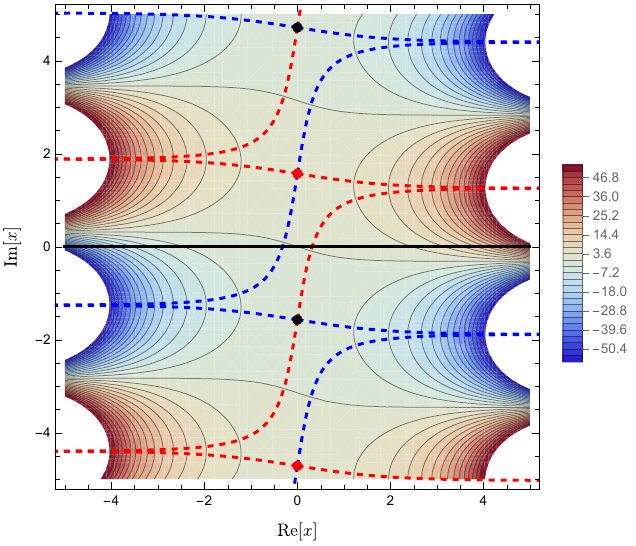}}
\caption{Complex deformation of Fig.~\ref{fig:Picard-Lefschetz2} with
$\hbar = e^{i\theta }$. (a) $\theta = +\frac{\pi}{10}$, (b) $\theta = -\frac{\pi}{10}$. 
}
\label{fig:Picard-Lefschetz3}
\end{figure} 
%
%

On the other hand, in the case where the parameters $\beta$ and $\gamma$ have opposite signs, the steepest-descent contours are shown in Fig.~\ref{fig:Picard-Lefschetz2}. 
We can find that the Lefschetz thimbles pass multiple saddle points and the thimble decomposition becomes ambiguous. 
To see the structures in more detail, we deform the parameter $\hbar$ to be complex as $\hbar =|\hbar |e^{i\theta}$, perform the Lefschetz thimble analysis for complex $\hbar$ and take the limits $\theta \rightarrow 0_\pm$ to approach real positive $\hbar$ as in \cite{Honda:2024aro}.%
\footnote{
To ensure the convergence of the integral at $|x|\rightarrow \infty$, we shift the integral contour to imaginary direction as $\mathbb{R}-i\epsilon$ satisfying $\epsilon > |\theta |$.  
Therefore, if we take e.g.~$\epsilon =\mathcal{O}(\theta )$ s.t. $\epsilon > |\theta |$, then it gives a natural deformation of $G(\hbar )$ to complex $\hbar$. 
} 
After the complex deformation, the contour plots of $\text{Re}[F_D(x)]$ are illustrated in Fig. \ref{fig:Picard-Lefschetz3}. From this figure, it is evident that the contributing saddle is the same irrespective of the sign of the deformation parameter $\theta$. Specifically, the saddle is given by 
$x^{(0)}_{s+}= \frac{1}{2} \log \left| \frac{\gamma}{\beta} \right| -\frac{\pi i}{2}$.%
\footnote{
When $\gamma /\beta <0$, identification of the sign $\pm$ of the saddle point $x_{s\pm}^{(m)}$ in \eqref{eq:saddles} depends on how to take the branch cut.
Here we take $\sqrt{\gamma }=e^{-\frac{\pi i}{2}}\sqrt{|\gamma |}$ for $\gamma  <0$.
This is because, if we continuously change ${\rm arg}(\gamma )$ to connect $\gamma >0$ and $\gamma <0$, then there is a region in ${\rm arg}(\gamma ) \in [0,\pi]$ where the integral of $G(\hbar )$ becomes divergent while we can keep it well-defined in the whole region of ${\rm arg}(\gamma ) \in [-\pi ,0]$.  
Indeed Fig.~\ref{fig:Picard-Lefschetz2} can be obtained by continuously changing ${\rm arg}(\gamma )$ in the region ${\rm arg}(\gamma ) \in [-\pi ,0]$ from Fig.~\ref{fig:Picard-Lefschetz1} and vice versa.
}
Therefore, even though the parameters in Fig.~\ref{fig:Picard-Lefschetz2} are located on the Stokes lines before the deformation, the Stokes jump
does not occur. 
As in the previous case, 
changing the integral variable as $x=x_s+\delta x$,
we can perform the path integral,
\begin{align}
    G(\hbar) &=\frac{1}{4\hbar}\int_{-\infty}^\infty  \! \mathrm{d}\delta x \exp \left[-\frac{2}{\hbar}\sqrt{ |\beta\gamma| } \cosh ({\delta x})\right]\notag \\
    =& \frac{1}{2\hbar }K_0\left(\frac{2\sqrt{ |\beta\gamma| }}{\hbar }\right)
= - \frac{\pi i}{4\hbar }H^{(2)}_0\left( -\frac{2i\sqrt{|\beta\gamma | }}{\hbar }\right)\,,
\label{eq:G_K0r}
\end{align}
The asymptotic form of $G(\hbar) $ with $\hbar \to 0$ leads to
\begin{equation}
    G(\hbar) \simeq \frac{\pi^{\frac{1}{2}}}{4\hbar^{\frac{1}{2}} |\beta\gamma|^{\frac{1}{4}}}
    e^{-\frac{2}{\hbar}\sqrt{ |\beta\gamma| }}\,.
\end{equation}
The probability amplitude is exponentially suppressed and describes the quantum evolution from $(q_0,\varphi_0)$ to $(q_1,\varphi_1)$. In the scenario where $q_0 = 0$ and $\varphi_0 > \varphi_1$, these probability amplitudes correctly describe the two-dimensional quantum creation of the universe.
By using $x^{(0)}_{s+}$, the saddle points in terms of the lapse function lie in the Euclidean direction, with $\textrm{Im}[N^{(0)}_{s+}] < 0$,
indicating that the original version of the no-boundary proposal is approximately correct for \ac{jt} gravity. This is in contrast to the four-dimensional case~\cite{Honda:2024aro}.
Depending on the sign of the parameters $\beta$ and $\gamma$, the saddle points $x_{s}$ can be real or complex, but in both cases, the probability amplitudes are written in terms of the modified Bessel function of the second kind $K_0(z)$ when the Dirichlet boundary condition is employed. In the next section, we will consider the \ac{jt} probability amplitudes for Neumann boundary conditions.

\section{Neumann boundary condition for JT gravity}
\label{sec:Neumann-boundary-condition}

The Dirichlet boundary condition can be used to set the size of the universe to zero on the initial hypersurface, capturing the concept of the quantum creation of the universe from nothing. However, other boundary conditions such as Neumann, and Robin boundary conditions
are possible in Lorentzian path integral~\cite{DiTucci:2019dji,DiTucci:2019bui,Narain:2021bff,
Narain:2022msz,Ailiga:2023wzl,Ailiga:2024mmt}. In this section, we will consider the Neumann boundary condition for the scale factor and the dilaton field on the initial hypersurfaces.

We start with the following \ac{jt} action, 
\begin{align}
S_{\rm JT}=-2\pi\int\mathrm{d}t \left(\frac{\dot{q}\dot{\varphi}}{4N}
+N\Lambda \right)+S_B \,,
\end{align}
where we introduced the boundary term. 
Since the \ac{jt} action $S_{\rm JT}$ depends on $q,\dot{q},\varphi,\dot{\varphi}$, 
the variation of the action is given by
\begin{align}
\delta S_{\rm JT}&=-2\pi\int\mathrm{d}t \left(\left[-\frac{\ddot\varphi}{4N}\right]\delta q +\left[-\frac{\ddot q}{N}\right]\delta \varphi\right)\notag  \\
&-\frac{\pi}{2N}\dot\varphi\delta q -\frac{\pi}{2N}\dot q\delta \varphi+\delta S_B \,,
\end{align}
To derive the equation of motion, we can choose several boundary conditions for the scale factor and the dilaton field, including Dirichlet, Neumann, and Robin conditions, as well as specific boundary terms $S_B$ located on the hypersurfaces at $t_{i,f} = 0, 1$. To apply the Neumann boundary condition for the scale factor and the dilaton field on the initial hypersurfaces, we assume the following boundary term,
\begin{align}
S_B= \frac{1}{2}\int_{\partial \mathcal{M}}\mathrm{d}y  \sqrt{\gamma}
\left(\frac{\dot \phi}{N} + \phi\, \mathcal{K}\right)
=\frac{\pi}{2N}\left(\dot q\varphi+ q \dot \varphi\right)_{|_{t_i=0}}\,,
\end{align}
where 
we fix the value of the scale factor and dilaton field at the two endpoints,
\begin{align}\label{eq:Neumann-boundary}
\begin{split}
-&\frac{\pi\dot q(t_{i}=0)}{2N}=\Pi_{\varphi}^{0}, 
\ -\frac{\pi\dot \varphi(t_{i}=0)}{2N}=\Pi_{q}^{0}, \\
& q(t_{f}=1)=q_{1}, \ 
\varphi(t_{f}=1)=\varphi_{1}\,,    
\end{split}
\end{align} 
where the canonical momenta are defined as $\Pi_{q,\varphi}= \frac{
\partial L_{\rm JT}}{\partial \dot{q},\dot{\varphi}}$ 
in which $L_{\rm JT}$ is the \ac{jt} Lagrangian.
In Lorentzian path integral of the Kantowski-Sachs model, Refs.~\cite{Fanaras:2021awm, Fanaras:2022twv} assumes different boundary term and condition following the no-boundary proposal and tunneling proposal. Now, we consider the simple boundary terms and conditions.
For the above boundary condition~\eqref{eq:Neumann-boundary}, the path integral 
can be exactly evaluated (see Appendix~\ref{sec:appendix-prefactor} for the detail derivation). On the other hand, by using the initial momentum state, 
\begin{equation}
\left|\Pi_{q}^{0},\Pi_{\varphi}^{0}\right\rangle
=\frac{1}{2\pi \hbar} \int_{-\infty}^{\infty} \mathrm{d}q_{0}\mathrm{d}\varphi_{0} e^{\frac{i}{\hbar}
(\Pi_{q}^{0}q_{0}+\Pi_{\varphi}^{0}\varphi_{0})}\left|q_{0},\varphi_{0}\right\rangle\,,
\end{equation}
we can also and easily derive the probability amplitude for the Neumann boundary condition as a Fourier transform of another probability amplitude~\cite{Ailiga:2023wzl},
\begin{align}
\begin{split}
&G[q_{1},\varphi_{1};\Pi_{q}^{0},\Pi_{\varphi}^{0}]=\\
&\frac{1}{2\pi \hbar} \int_{-\infty}^{\infty} \mathrm{d}q_{0}\mathrm{d}\varphi_{0} e^{i (\Pi_{q}^{0}q_{0}+\Pi_{\varphi}^{0}\varphi_{0})/\hbar}
G[q_{1},\varphi_{1};q_{0},\varphi_{0}]\,.   
\end{split}
\end{align}
Thus, we can obtain
\begin{align}\label{G-amplitude-Neumann}
&G[q_{1},\varphi_{1};\Pi_{q}^{0},\Pi_{\varphi}^{0}]
= \nonumber \\
&\frac{1}{2\pi \hbar} \int\! \mathrm{d}N e^{\frac{i}{\hbar}\left\{
-2\pi\Lambda N + \frac{2 N \Pi_{q}^{0}\Pi_{\varphi}^{0}}{\pi} 
+ q_1 \Pi_{q}^{0} + \varphi_1 \Pi_{\varphi}^{0}\right\}}\,,
\end{align}
where in the lapse integration, the absence of a pole means 
there is no ambiguity in the integration contours.
Now, we will consider the integration range of $N \in (0, \infty)$. 
The above integration can be performed for $\textrm{Im}[\hbar]<0$ 
and we can obtain
\begin{equation}
G[q_{1},\varphi_{1};\Pi_{q}^{0},\Pi_{\varphi}^{0}]= -\frac{i e^{\frac{i}{\hbar }(\Pi_{q}^{0}q_1 + \Pi_{\varphi}^{0} 
\varphi_1)} }{4(\pi^2\Lambda - \Pi_{q}^{0}\Pi_{\varphi}^{0})}\,.  
\end{equation}
While the probability amplitude under Dirichlet boundary conditions corresponds to the solution~\eqref{eq:wdw-minisuperspace-solution1} and~\eqref{eq:wdw-minisuperspace-solution2}, the amplitude under Neumann boundary conditions does not necessarily do so. When $\Pi_{q}^{0},\Pi_{\varphi}^{0}$ are real, the probability amplitudes oscillate, representing the classical evolution of the universe. In contrast, when $\Pi_{q}^{0},\Pi_{\varphi}^{0}$ are imaginary, the probability amplitudes become exponentially suppressed, corresponding to the quantum genesis of the universe.

\section{Quantum perturbation problems}
\label{sec:perturbation-problems}

In the previous sections, we have calculated the probability amplitude of the 
scalar factor and dilaton field based on the framework of Lorentzian quantum cosmology and \ac{jt} gravity.
However, since we consider cosmological genesis, it is necessary to address not only the background spacetime and homogeneous dilaton field but also perturbations as a practical problem. In Lorentzian quantum cosmology, the issue of spacetime perturbations has been extensively discussed so far~\cite{Feldbrugge:2017fcc,DiazDorronsoro:2017hti, Feldbrugge:2017mbc, Feldbrugge:2018gin, DiazDorronsoro:2018wro, Halliwell:2018ejl, Janssen:2019sex, Vilenkin:2018dch, Vilenkin:2018oja, Bojowald:2018gdt, DiTucci:2018fdg, DiTucci:2019dji, DiTucci:2019bui, Lehners:2021jmv, Matsui:2022lfj, Matsui:2024bfn,Ailiga:2024nkz}. Now, we shall discuss the problem of perturbations in quantum cosmology within the \ac{jt} gravity.
Since the gravitational wave does not propagate in \ac{jt} gravity, instead we shall consider a massless scalar field as the perturbations around the background, 
\begin{align}
 S_{\Phi} &= -\frac{1}{2}\int_{\mathcal{M}}\mathrm{d}^2 x \sqrt{-g}
[g^{\mu\nu}\nabla_{\mu} \Phi\nabla_{\nu}\Phi] \notag \\
&= -\frac{1}{2}\int \mathrm{d}t\mathrm{d}x Na
[-\frac{1}{N^2}\dot{\Phi}^2 + \frac{1}{a^2}{\Phi'}^2]\,.
\label{eq:action-perturbation}
\end{align}
We expand the real scalar field $\Phi$ as 
\begin{equation}
\Phi(t,x) =\frac{1}{\sqrt{2\pi}}\sum_{n=-\infty}^\infty \Phi_n(t) e^{in x}\,,
\end{equation}
where $\Phi_n(t)=\Phi_{-n}^*(t)$.
By using the above expression, 
we obtain 
\begin{align}\label{eq:action-perturbation-mode}
 S_{\Phi} = \frac{1}{2}\int \mathrm{d}t
N\sum_{n=-\infty}^\infty \Bigl[ \frac{q\varphi^{\frac{1}{2}}}{N^2}\dot{\Phi}_n\dot{\Phi}_{-n} - \frac{n^2}{q\varphi^{\frac{1}{2}}}{\Phi}_n{\Phi}_{-n} \Bigr]\,.
\end{align}
where we transformed $N\to N/a\phi$, and
defined $q(t)=a^2(t)$ and $\varphi(t)=\phi^2(t)$.

If the backreaction of the perturbation on the background can be neglected, the path integral is first evaluated for $q$ and $\varphi$, and then the on-shell action of the perturbation is computed by finding the classical solution of the perturbation, utilizing the classical solutions of $q$ and $\varphi$. The total on-shell action is given by $S_{\textrm{on-shell}}[N]+S_{\textrm{on-shell},{\Phi}}[N]$
and can be evaluated approximately by the saddle-point method of $N$ integral.
The probability amplitude of \ac{jt} gravity, taking into account perturbations, is given by
\begin{align}\label{G-total-amplitude}
&G\left[q_{1},\varphi_{1},\Phi_{n}\right]\nonumber \\
&\quad
= \int_{0}^\infty \! \frac{\mathrm{d}N }{4\hbar N\cdot D(N)}
e^{\frac{i}{\hbar}(
S_\textrm{on-shell}[N]+S_{\textrm{on-shell},{\Phi}}[N])}\,,
\end{align}
where $D(N)$ is the the functional
determinant of $\Phi_{n}$. Here, we do not evaluate functional determinants of perturbations, but they can be analyzed in principle. The above evaluation can be expected to provide a sufficiently accurate analysis.

For simplicity, we shall focus on the one-mode scalar field ${\Phi}_n$.
The equation of motion is given by
\begin{equation}
\frac{1}{N^2}\ddot{\Phi}_n+\frac{1}{N^2}\frac{\dot{\chi}}{\chi}\dot{\Phi}_n
+\frac{n^2}{\chi^2}{\Phi}_n=0\,,
\end{equation}
where $\chi=q\varphi^{\frac{1}{2}}$.
By using the solutions $q(t)=(q_1 -q_0 )t+q_0$ and 
$\varphi(t)=(\varphi_1 -\varphi_0 )t+\varphi_0$ and setting $q_0=0$ for simplicity, 
we obtain 
\begin{align}\label{eq:perturbative-solution}
\begin{split}
{\Phi}_n(t) &=  C_1 e^{-\frac{2 n N \tan ^{-1}\left(\frac{\sqrt{(t-1) \varphi _0-t \varphi _1}}{\sqrt{\varphi _0}}\right)}{q_1 \sqrt{\varphi _0}}} \\
&\quad + C_2 e^{+\frac{2 n N \tan ^{-1}\left(\frac{\sqrt{(t-1) \varphi _0-t \varphi _1}}{\sqrt{\varphi _0}}\right)}{q_1 \sqrt{\varphi _0}}}\,.   
\end{split}
\end{align}

Now, we shall impose the following boundary conditions, $\lim_{\, \epsilon\to 0}{\Phi}_n(t=\epsilon)=0$
and ${\Phi}_n(t=1)={\Phi}_{n1}$ since it is reasonable to assume the perturbation is sufficiently small as an initial condition, and we fix the coefficients $C_{1,2}$.
By performing the integration by parts for the action~\eqref{eq:action-perturbation-mode} and using the equation of motion for ${\Phi}_n(t)$, we can get the on-shell action, 
\begin{equation}\label{eq:on-shell-action-perturbation}
S_{\textrm{on-shell},{\Phi}_n}[N] = 
\left[ \frac{q\varphi^{\frac{1}{2}}\dot{\Phi}_n{\Phi}_{-n}}{2N} \right]^1_0\,.
\end{equation}
Inserting the solution~\eqref{eq:perturbative-solution} into the on-shell action~\eqref{eq:on-shell-action-perturbation}, we can get 
\begin{align}
\begin{split}
&S_{\textrm{on-shell},{\Phi}_n}[N]  = \lim_{\epsilon\to 0}
\frac{n \Phi_1^2 }{2 i}\times \\
& \coth \left(\frac{2 n N \left(-\tan ^{-1}\left(\frac{\sqrt{\varphi _0 (\epsilon -1)-\varphi _1 \epsilon }}{\sqrt{\varphi _0}}\right)+\tan ^{-1}\left(\frac{i \sqrt{\varphi _1}}{\sqrt{\varphi _0}}\right)\right)}{q_1 \sqrt{\varphi _0}}\right)\,.    
\end{split}
\end{align}
For the complex saddle point $N_s$, we obtain the following expression, 
\begin{equation}
S_{\textrm{on-shell},{\Phi}_n}[N_s]  = \left\{
\begin{array}{l}
\vspace{0.2cm}
-\frac{in}{2}{\Phi}_{n1}^2\quad \left(\textrm{Im}[N_s]>0\right)\,,\\
\vspace{0.2cm}
+\frac{in}{2}{\Phi}_{n1}^2\quad \left(\textrm{Im}[N_s]<0\right)\,,
\end{array}
\right. 
\end{equation}
For $\textrm{Im}[N_s] > 0$, the on-shell action of the perturbation leads to $\textrm{Re}[iS_{\textrm{on-shell}, \Phi_n}[N_s]/\hbar] > 0$, implying that the probability amplitude of the perturbation is inverse-Gaussian, which suggest the perturbation around background is not significantly suppressed. On the other hand, for $\textrm{Im}[N_s] < 0$, we can obtain a Gaussian probability amplitude for the perturbation and the scalar field perturbation is a ground state.
In the case where $q_0 = 0$ and $\varphi_0 > \varphi_1$, the dominant saddle points lie in the Euclidean direction, where $\textrm{Im}[N_s] < 0$,
and the probability amplitudes accurately describe the two-dimensional quantum genesis of the universe. The original formulation of the no-boundary proposal remains approximately valid and perturbative regularity is preserved. This result contrasts with that in the four-dimensional GR case, where the regularity is disrupted.

On the other hand, for the real saddle point $N_s$ setting $q_0=\varphi_0=0$, we get 
\begin{align}
&S_{\textrm{on-shell},{\Phi}_n}[N_s]= \lim_{\epsilon\to 0}\frac{in}{2}{\Phi}_{n1}^2 \nonumber \\
& \times \frac{\left(1+
\cos {\frac{4 n N_s }{q_1\sqrt{\varphi_1}}\left(1-\frac{1}{\sqrt{\epsilon }}\right)}
+i \sin {\frac{4n N_s }{q_1\sqrt{\varphi_1}}\left(1-\frac{1}{\sqrt{\epsilon }}\right)}
\right)}{\left(1-
\cos {\frac{4 n N_s }{q_1\sqrt{\varphi_1}}\left(1-\frac{1}{\sqrt{\epsilon }}\right)}
-i \sin {\frac{4n N_s }{q_1\sqrt{\varphi_1}}\left(1-\frac{1}{\sqrt{\epsilon }}\right)}
\right)}\,.
\end{align}
In the scenario where $q_0 = \varphi_0 = 0$, 
the dominant saddle points lie in the real direction, indicating that the perturbative on-shell action $S_{\textrm{on-shell},{\Phi}_n}[N]$ can not be clearly defined and the probability amplitude of \ac{jt} gravity is not well-defined.

\section{Conclusions}
\label{sec:conclusions}
In this paper, we have studied the probability amplitude of \ac{ds} spacetime in \ac{jt} gravity. Our approach uses the Lorentzian path integral formalism, which differs from previous studies that primarily rely on solutions of the \ac{wdw} equation or path integrals with Euclidean metrics. Instead, we have directly computed the \ac{jt} probability amplitudes via the Lorentzian path integral, incorporating both determinant factors and physical boundary conditions. In particular, we focus on the Dirichlet boundary conditions for the scale factor and the dilaton field.

We have shown that under Dirichlet boundary conditions, the amplitude derived from the Lorentzian path integral takes a form expressed in terms of the modified Bessel function of the second kind, $K_0(z)$. Although the derived amplitude is consistent with the solutions of the \ac{wdw} equation, previous works~\cite{Maldacena:2019cbz,Iliesiu:2020zld} left ambiguity as to which Bessel function is appropriate to represent the \ac{jt} wave function. Once the boundary conditions for the scale factor and the dilaton field are fixed, the Lorentzian path integral can determine a unique \ac{jt}  probability amplitude without contradiction.
In addition, we provide a detailed analysis of the thimble structure associated with the path integral of \ac{jt} gravity, carefully examining the relevant saddle points and their steepest descent contours to evaluate the quantum amplitudes accurately. 
The corresponding saddle points $N_s$ lie on the real axis or the complex plane of the lapse function where $\textrm{Im}[N_s] < 0$. Therefore, our results suggest that the no-boundary proposal is approximately valid in the Lorentzian quantum cosmology of \ac{jt} gravity, in contrast to four-dimensional Einstein gravity.

Furthermore, we have shown that a stable quantum genesis of the universe occurs only when the initial value of the dilaton field is non-zero and large. 
In other words, when the size of the universe approaches zero, the value of the dilaton field cannot be set to zero. This result implies that certain initial conditions, particularly a non-zero dilaton field, are necessary for a perturbatively stable universe to emerge from nothing in \ac{jt} gravity.

In this study, we successfully evaluated the \ac{jt} probability amplitude using the Lorentzian path integral with minisuperspace approximation. However, as shown in Section~\ref{sec:WdW-wavefunctional}, it is also possible to find a general solution to the \ac{wdw} equation without relying on this approximation. We expect that the amplitude can be accurately evaluated beyond the minisuperspace by employing the Lorentzian path integral, and we plan to explore this extension in future work.

Additionally, \ac{jt} gravity has been widely discussed in relation to holography, particularly in the context of the \ac{ds}/CFT correspondence. It would be valuable to investigate the holographic dual of our findings to test this correspondence. We will consider this research in future investigations.

\bigskip
\section*{Acknowledgement}

We would like to thank Takahiro Terada for useful comments and for carefully checking our manuscript.
M.~H.~is supported by JST PRESTO Grant Number JPMJPR2117, JST CREST Grant Number JPMJCR24I3 and JSPS Grant-in-Aid for Transformative Research Areas (A) ``Extreme Universe" JP21H05190 [D01].
This work is supported by Japan Society for the Promotion of Science (JSPS) KAKENHI Grant Numbers JP22H01222(M.H.), JP22KJ1782(H.M.), JP23K13100(H.M.), JP23KJ1090 (K.N.), and JP23KJ1162 (K.O.).

\onecolumngrid

\appendix

\section{Computation of prefactor}
\label{sec:appendix-prefactor}

In this appendix, we explicitly calculate the Lorentzian transition amplitude \eqref{Lorentzian-amplitude} under 
\begin{enumerate*}
    \item Dirichlet-Dirichlet 
    \item Neumann-Neumann
    \item Neumann-Dirichlet and Dirichlet-Neumann
\end{enumerate*}
boundary conditions.
The minisuperspace action is given as
\begin{align}
    S_{\rm JT}
    &=-2\pi\int^{t_f}_{t_i} \mathrm{d}t
    \left(\frac{1}{4N} \frac{dq}{dt} \frac{d\varphi}{dt}
    +N\Lambda \right)
    +S_B \nonumber\\
    &=-2\pi\int^{1}_{0} \mathrm{d}\tau
    \left(\frac{\dot{q}\dot{\varphi}}{4\tilde{N}}
    +\tilde{N} \Lambda \right)
    +S_B\, ,
    \label{eq:action-jt-minisuperspace_re}
\end{align}
where $S_B$ is the boundary term other than the Gibbons-Hawking-York term to realize the boundary conditions. In the second line, we change the integration variable as $t\rightarrow \tau= (t-t_i)/(t_f-t_i)$ with $\tilde{N}=N(t_f-t_i)$. The dot denotes the derivation with respect to $\tau$ throughout this appendix. What we are interested in is the following part of the transition amplitude
\begin{align}
    P (N) 
    := \int \mathcal{D}q\mathcal{D}\varphi 
    \exp\left(iS_{\rm JT}[N,q,\varphi]/\hbar\right)\,.
\end{align}
We calculate this amplitude with the proper normalization prefactor under the mentioned boundary conditions. In the following, we omit the contribution from the cosmological constant in \eqref{eq:action-jt-minisuperspace_re} which is almost trivial, and focus on the contribution from the gravitational action with boundary terms. That is
\begin{align}
P_{q\varphi} (N) 
:= \int \mathcal{D}q\mathcal{D}\varphi 
\exp\left((iS_{q\varphi}+S_B)/\hbar\right)\,,
\label{eq:rel_TA}
\end{align}
with 
\begin{align}
S_{q\varphi}
:=-\frac{\pi}{2\tilde{N}} \int_0^1 \mathrm{d}\tau \left( \dot{q}\dot{\varphi}  \right)\,.
\label{eq:grav_action}
\end{align}
As mentioned in the footnote around \eqref{Lorentzian-amplitude}, we note that $q$ and $\varphi$ are extended to include negative values. This extension avoids the issues that arise from the non-unique classical paths, which can occur if the variables are restricted to be positive \cite{Halliwell:1988wc}. The consequences are not altered if we choose the boundary conditions consistent with their original positivities.

\subsection{Dirichlet-Dirichlet}
Here we consider the case with the following Dirichlet-Dirichlet type boundary conditions:
\begin{equation}
q(\tau=0)=q_{0}, \ \varphi(0)=\varphi_{0}, \ q(1)=q_{1}, \ 
\varphi(1)=\varphi_{1}.
\end{equation} 
We note that the boundary action is not necessary here
\begin{align*}
    S_B = 0\, .
\end{align*}

To compute the prefactor, it is convenient to redefine the fields as
\begin{equation}
\tilde{q} (\tau) := q(\tau) -q_0 -(q_1 -q_0 )\tau ,\quad
\tilde{\varphi} (\tau) := \varphi (\tau) -\varphi_0 -(\varphi_1 -\varphi_0 )\tau ,
\end{equation}
which satisfy the boundary conditions
\begin{equation}
\tilde{q} (0)=
\tilde{q} (1) = 0 ,\quad
\tilde{\varphi} (0)=\tilde{\varphi} (1) = 0.
\label{eq:BC_DD}
\end{equation}
Note that this redefinition does not give a nontrivial Jacobian factor in the path integral.
Additionally, let us consider the mode expansions
\begin{equation}
    \tilde{q}(\tau) =\sum_{n=1}^\infty \tilde{q}_n \sin\Big[\pi n \tau\Big] ,\quad
    \tilde{\varphi}(\tau)= \sum_{n=1}^\infty \tilde{\varphi}_n \sin\Big[\pi n \tau\Big] \, .
\end{equation}
These modes obviously satisfy the boundary conditions \eqref{eq:BC_DD}. The real condition of $q, \varphi$ demands real coefficients $\{\tilde{q}_n\}, \{\tilde{\varphi}_n\}$

Then, the gravitational action \eqref{eq:grav_action} becomes
\begin{eqnarray}
    S_{q\varphi}
    &=& -\frac{\pi}{2\tilde{N}} \int_0^1 \mathrm{d}\tau  \Bigl[ 
    \left( \dot{\tilde{q}} +(q_1 -q_0 )  \right)  
    \left( \dot{\tilde{\varphi}}  +(\varphi_1 -\varphi_0 )  \right)\Bigr] \nonumber \\
    &=& -\frac{\pi^3}{4\tilde{N}} \sum_{n=1}^\infty n^2  
    \tilde{q}_n\tilde{\varphi}_{n}
    -\frac{\pi}{2\tilde{N}} (q_1 -q_0 )(\varphi_1 -\varphi_0 ) .
\end{eqnarray}
The amplitude \eqref{eq:rel_TA} reduces to
\begin{eqnarray}
    P_{q\varphi} (N) 
    &=& C_{DD}(t_f,t_i) 
    \exp\Bigg[
        -\frac{i\pi}{2\hbar \tilde{N}} 
        (q_1-q_0)(\varphi_1-\varphi_0)
    \Bigg]
    \int  \left( \prod_{n=1}^\infty d\tilde{q}_n~d\tilde{\varphi}_n  \right)  
    \exp\Bigg[ 
        -\frac{i\pi^3}{4\hbar\tilde{N}} 
        \sum_{n=1}^\infty n^2  
        \tilde{q}_n\tilde{\varphi}_{n}
    \Bigg] 
    \nonumber \\ 
    &=& C_{DD}(t_f,t_i) 
    \exp\Bigg[
        -\frac{i\pi}{2\hbar \tilde{N}} 
        (q_1-q_0)(\varphi_1-\varphi_0)
    \Bigg]
    \prod_{n=1}^\infty 
    \Bigg[ 2\pi \int  d\tilde{\varphi}_n \ 
    \delta\Bigg( 
        \frac{\pi^3}{4\hbar\tilde{N}} 
        n^2  \tilde{\varphi}_{n}
    \Bigg) 
    \Bigg]
    \nonumber \\
    &=& C_{DD}(t_f,t_i) 
    \exp\Bigg[
        -\frac{i\pi}{2\hbar \tilde{N}} 
        (q_1-q_0)(\varphi_1-\varphi_0)
    \Bigg]
    \prod_{n=1}^\infty \Biggl[ \frac{8\hbar \tilde{N}}{\pi^2 n^2} \Biggr] ,
\end{eqnarray}
with some normalization constant $C_{DD}(t_f,t_i)$. Applying the zeta functional regularization
\begin{eqnarray}
    \prod_{n=1}^\infty c 
    &=& \exp{\left( \sum_{n=1}^\infty \log{c} \right) }
    = \exp{\left( \zeta (0) \log{c}  \right) }
    = \exp{\left( -\frac{1}{2} \log{c}  \right) } = c^{-1/2} , \nonumber \\
    \prod_{n=1}^\infty \frac{1}{n^2} 
    &=& \exp{\left( -2\sum_{n=1}^\infty  \log{n} \right) }
    = \exp{\left( 2\zeta'(0)  \right) } 
    = \exp{\left( -\log{(2\pi)} \right) } 
    =\frac{1}{2\pi} ,
    \label{eq:zeta_reg}
\end{eqnarray}
we obtain
\begin{eqnarray}
    P_{q\varphi}(N;\;q_1, \varphi_1; \;q_0, \varphi_0) 
    &=& \frac{C_{DD}(t_f,t_i) }{ \sqrt{32\hbar N(t_f-t_i)} }\exp\Bigg[
        -\frac{i\pi}{2\hbar N(t_f-t_i)} 
        (q_1-q_0)(\varphi_1-\varphi_0)
     \Bigg]\, .
\label{eq:TA_DD}
\end{eqnarray}

The normalization condition is obtained from the relation
\begin{align}
    \int \mathrm{d}q \mathrm{d}\varphi P_{q\varphi}(N;\; q_1, \varphi_1; \;q, \varphi) 
    P_{q\varphi}(N;\; q, \varphi; \;q_0, \varphi_0) 
    = P_{q\varphi}(N;\; q_1, \varphi_1; \;q_0, \varphi_0) \, .
\end{align}
Inserting \eqref{eq:TA_DD} into the above relation, the left-hand side gives
\begin{align}
    (\mathrm{L.H.S})
    &= \Biggl(\frac{1}{ \sqrt{32\hbar\tilde{N}}} \Biggr)^2
    \frac{C_{DD}(t_f,t)C_{DD}(t,t_i)  }{\sqrt{(t_f-t)(t-t_i)}}
    \int \mathrm{d}q \mathrm{d}\varphi 
    \exp\Bigg[
    -\frac{i\pi}{2\hbar N} 
    \Bigg[
        \frac{(q_1 -q )(\varphi_1 -\varphi)}{t_f-t}
        + \frac{(q -q_0 )(\varphi -\varphi_0 )}{(t-t_i)}
    \Bigg]
    \Bigg] 
    \nonumber \\
     &= \Biggl(\frac{1}{ \sqrt{32\hbar\tilde{N}}} \Biggr)^2
    \frac{C_{DD}(t_f,t)C_{DD}(t,t_i)  }{\sqrt{(t_f-t)(t-t_i)}}
     \exp\Bigg[
     -\frac{i\pi}{2\hbar N} 
     \Bigg(\frac{q_1 \varphi_1}{t_f-t} + \frac{q_0 \varphi_0}{t-t_i}
     \Bigg)
     \Bigg]
     \nonumber \\
     & \hspace{80pt} \cdot 2\pi
     \int \mathrm{d}\varphi \  
     \delta \Bigg[
        \frac{\pi}{2\hbar N}    
        \Bigg(
        \frac{\varphi-\varphi_1}{t_f-t}
        +\frac{\varphi-\varphi_0}{t-t_i}
        \Bigg)
     \Bigg]
     \exp\Bigg[
     \frac{i\pi}{2\hbar N} 
        \Bigg(
        \frac{q_1}{t_f-t} + \frac{q_0}{t-t_i}
        \Bigg)\varphi
     \Bigg] \nonumber \\
     &= 4 \hbar N C_{DD}(t_f,t)C_{DD}(t,t_i)
     \Biggl(\frac{1}{ \sqrt{32\hbar\tilde{N}}} \Biggr)^2
    \frac{\sqrt{(t_f-t)(t-t_i)}}{t_f-t_i}
     \exp\Bigg[
     -\frac{i\pi}{2\hbar N(t_f-t_i)} 
     (q_1-q_0)(\varphi_1-\varphi_0)
     \Bigg]\, ,
\end{align}
It is noticed that the consistent normalization factor is found as
\begin{align}
    C_{DD}(t_f,t_i)  = \sqrt{\frac{2}{\hbar N (t_f-t_i)}}\, ,
\end{align}
and the whole normalized transition amplitude \eqref{eq:action-jt-minisuperspace_re} with Dirichlet-Dirichlet type boundary condition results in
\begin{align}
    P(N;\; q_1, \varphi_1; \;q_0, \varphi_0) 
    = \frac{1}{ 4\hbar N (t_f-t_i)} 
    \exp\Bigg[
    -\frac{i\pi}{\hbar}
    \Bigg(
     \frac{1}{2N(t_f-t_i)} 
     (q_1-q_0)(\varphi_1-\varphi_0)
     + 2N \Lambda (t_f-t_i)
    \Bigg)
     \Bigg]\, .
    \label{eq:TA_DD_nom}
\end{align}

Compared to the 4-D GR solution \cite{Feldbrugge:2017kzv}, the main difference is that the prefactor in 2-D JT gravity case is proportional to $N^{-1}$ rather than $N^{-1/2}$. This stems from the difference in the number of independent variables. On the other hand, the phase factor looks consistent, taking the conformal coupling and the dimensionality into account.

\subsection{Neumann-Neumann}
Next, we consider the case with the following Neumann-Neumann boundary conditions:
\begin{equation}
    -\frac{\pi}{2\tilde{N}}\dot{q}(0)=\Pi_{\varphi}^{0}, \quad
    -\frac{\pi}{2\tilde{N}}\dot{\varphi}(0)=\Pi_{q}^{0}, \quad
    -\frac{\pi}{2\tilde{N}}\dot{q}(1)=\Pi_{\varphi}^{1}, \quad
    -\frac{\pi}{2\tilde{N}}\dot{\varphi}(1)=\Pi_{q}^{1}\, .
\end{equation} 
where $\tilde{N}= N (t_f-t_i)$. This condition is not realized as a classical solution unless $\Pi_{\varphi}^{0} = \Pi_{\varphi}^{1}$ and $\Pi_{q}^{0}= \Pi_{q}^{1}$. Indeed, the viable solution is only this coincident case even at the quantum level, as we will show later. The boundary term for this condition turns out to be
\begin{align}
 S_B
:=-\Big(
\Pi_{\varphi}^{1} \varphi(1)+q(1) \Pi_{q}^{1}
-\Pi_{\varphi}^{0} \varphi(0)-q(0) \Pi_{q}^{0}
\Big)
\,.
\end{align}

As in the previous case, we redefine the fields as
\begin{equation}
\dot{\tilde{q}} (\tau) := \dot{q}(\tau) +\frac{2\tilde{N}}{\pi} \Pi_{\varphi}^{0} 
+\frac{2\tilde{N}}{\pi}(\Pi_{\varphi}^{1} -\Pi_{\varphi}^{0} )\tau,\quad
\dot{\tilde{\varphi}} (\tau) := \dot{\varphi}(\tau) +\frac{2\tilde{N}}{\pi} \Pi_{q}^{0} 
+\frac{2\tilde{N}}{\pi}(\Pi_{q}^{1} -\Pi_{q}^{0} )\tau ,
\end{equation}
or equivalently
\begin{equation}
\tilde{q} (\tau) := q(\tau) +\frac{2\tilde{N}}{\pi} \Pi_{\varphi}^{0} \tau
+\frac{\tilde{N}}{\pi}(\Pi_{\varphi}^{1} -\Pi_{\varphi}^{0} )\tau^2 ,\quad
\tilde{\varphi} (\tau) := \varphi(\tau) +\frac{2\tilde{N}}{\pi} \Pi_{q}^{0} \tau
+\frac{\tilde{N}}{\pi}(\Pi_{q}^{1} -\Pi_{q}^{0} )\tau^2  ,
\end{equation}
which satisfies the boundary conditions
\begin{equation}
\dot{\tilde{q}} (0)=
\dot{\tilde{q}} (1) = 0 ,\quad
\dot{\tilde{\varphi}} (0)= 
\dot{\tilde{\varphi}} (1) = 0 .
\end{equation}
Note that this redefinition does not give a nontrivial Jacobian factor in the path integral.

We perform the mode expansion as
\begin{equation}
\tilde{q}(\tau) =\sum_{n=0}^\infty \tilde{q}_n \cos\Big[\pi n \tau\Big] ,\quad
\tilde{\varphi}(\tau)= \sum_{n=0}^\infty \tilde{\varphi}_n \cos\Big[\pi n \tau\Big] ,
\end{equation}
One can see that each basis obviously satisfies the boundary conditions.
Then, the gravitational bulk action is reduced to
\begin{eqnarray}
S_{q\varphi}
&=& -\frac{\pi}{2\tilde{N}} \int_0^1 \mathrm{d}\tau  \Bigg[ 
\left( \dot{\tilde{q}} - \frac{2\tilde{N}}{\pi}(\Pi_{\varphi}^{1} -\Pi_{\varphi}^{0} )\tau -\frac{2\tilde{N}}{\pi} \Pi_{\varphi}^{0}\right)  
\left( \dot{\tilde{\varphi}}  -\frac{2\tilde{N}}{\pi}(\Pi_{q}^{1} -\Pi_{q}^{0} )\tau
-\frac{2\tilde{N}}{\pi} \Pi_{q}^{0}\right)\Bigg] \nonumber \\
&=& -\frac{\pi^3}{4 \tilde{N}} \sum_{n=1}^{\infty} n^2\left(\tilde{q}_n \tilde{\varphi}_n\right)
- (\Pi_{\varphi}^{1} -\Pi_{\varphi}^{0} ) \tilde{\varphi}_0
- (\Pi_{q}^{1} -\Pi_{q}^{0} ) \tilde{q}_0
-\frac{\tilde{N}}{3\pi} \Bigg[
\Pi_{\varphi}^{1} (2 \Pi_{q}^{1} + \Pi_{q}^{0})
+\Pi_{\varphi}^{0} (\Pi_{q}^{1} + 2\Pi_{q}^{0})
\Bigg]
\nonumber \\
&&\hspace{220pt}
+\Big(
\Pi_{\varphi}^{1} \tilde{\varphi}(1)+\tilde{q}(1) \Pi_{q}^{1}
-\Pi_{\varphi}^{0} \tilde{\varphi}(0)-\tilde{q}(0) \Pi_{q}^{0}
\Big).
\end{eqnarray}
Adding the boundary terms, the entire action is written as
\begin{align}
    S_{q\varphi} + S_{B}
    &= -\frac{\pi^3}{4 \tilde{N}} \sum_{n=1}^{\infty} n^2\left(\tilde{q}_n \tilde{\varphi}_n\right)
    - (\Pi_{\varphi}^{1} -\Pi_{\varphi}^{0} ) \tilde{\varphi}_0
    - (\Pi_{q}^{1} -\Pi_{q}^{0} ) \tilde{q}_0
    +\frac{\tilde{N}}{3\pi} \Bigg[
    \Pi_{\varphi}^{1} (\Pi_{q}^{1} + 2 \Pi_{q}^{0})
    +\Pi_{\varphi}^{0} (2 \Pi_{q}^{1} + \Pi_{q}^{0})
    \Bigg]\, .
\end{align}
Thus the amplitude \eqref{eq:rel_TA} is derived as
\begin{eqnarray}
P_{q\varphi} (N) 
&=& C_{NN} (t_f,t_i) \exp\Bigg[\frac{i\tilde{N}}{3\pi\hbar} \Big[
    \Pi_{\varphi}^{1} (\Pi_{q}^{1} + 2 \Pi_{q}^{0})
    +\Pi_{\varphi}^{0} (2 \Pi_{q}^{1} + \Pi_{q}^{0})
    \Big]
    \Bigg]
\nonumber \\
&& \hspace{20pt} 
\int  \left( \prod_{n=0}^\infty d\tilde{q}_n~d\tilde{\varphi}_n  \right) \exp\Bigl[ -\frac{i\pi^3}{4 \hbar \tilde{N}} \sum_{n=1}^{\infty} n^2\left(\tilde{q}_n \tilde{\varphi}_n\right)
- \frac{i}{\hbar}(\Pi_{\varphi}^{1} -\Pi_{\varphi}^{0} ) \tilde{\varphi}_0
- \frac{i}{\hbar}(\Pi_{q}^{1} -\Pi_{q}^{0} ) \tilde{q}_0\Bigr] 
\nonumber \\ 
&=& (2\pi)^2 C_{NN} (t_f,t_i) \exp\Bigg[\frac{2i\tilde{N}}{\pi\hbar}
    \Pi_{\varphi}^{1} \Pi_{q}^{1}
    \Bigg]
\delta \Bigg[\frac{\Pi_{\varphi}^{1} -\Pi_{\varphi}^{0} }{\hbar}\Bigg]
\delta \Bigg[\frac{\Pi_{q}^{1} -\Pi_{q}^{0} }{\hbar}\Bigg]
\prod_{n=1}^\infty \frac{8\hbar \tilde{N}}{\pi^2 n^2},
\end{eqnarray}
with normalization factor $C_{NN} (t_f,t_i)$. By applying the zeta functional regularization \eqref{eq:zeta_reg}, we obtain
\begin{eqnarray}
P_{q\varphi}(N;\; \Pi_{q}^{1}, \Pi_{\varphi}^{1}; \;\Pi_{q}^{0}, \Pi_{\varphi}^{0}) 
= \frac{C_{NN}(t_f,t_i)\pi^2 \hbar^{\frac{3}{2}}}{\sqrt{2N(t_f-t_i)}} 
\exp\Bigg[\frac{2iN(t_f-t_i)}{\pi\hbar}
\Pi_{\varphi}^{1} \Pi_{q}^{1}\Bigg]
\delta \Big[\Pi_{\varphi}^{1} -\Pi_{\varphi}^{0} \Big]
\delta \Big[\Pi_{q}^{1} -\Pi_{q}^{0} \Big]
\label{eq:TA_NN}
\end{eqnarray}

The normalization condition is obtained from the relation
\begin{align}
    \int \mathrm{d}p_q \mathrm{d}p_\varphi\; 
    P_{q\varphi}(N;\; \Pi_{q}^{1}, \Pi_{\varphi}^{1}; \;p_{q}, p_{\varphi})  
    P_{q\varphi}(N;\; p_{q}, p_{\varphi}; \;\Pi_{q}^{0}, \Pi_{\varphi}^{0})  
    = P_{q\varphi}(N;\; \Pi_{q}^{1}, \Pi_{\varphi}^{1}; \;\Pi_{q}^{0}, \Pi_{\varphi}^{0})  \, .
\end{align}
Inserting \eqref{eq:TA_NN} into above relation, it goes to
\begin{align}
    (\mathrm{L.H.S})
    &= \Biggl(\frac{\pi^2 \hbar^{\frac{3}{2}}}{\sqrt{2N}} \Biggr)^2
    \frac{C_{NN}(t_f,t)C_{NN}(t,t_i)}{\sqrt{(t_f-t)(t-t_i)}}
     \exp\Bigg[\frac{2iN(t_f-t_i)}{\pi\hbar}\Pi_{\varphi}^{1} \Pi_{q}^{1}\Bigg]
    \delta \Big[\Pi_{\varphi}^{1} -\Pi_{\varphi}^{0} \Big]
    \delta \Big[\Pi_{q}^{1} -\Pi_{q}^{0} \Big]
     \Bigg]\, .
\end{align}
The consistent normalization is
\begin{align}
    \frac{C_{NN}(t_f,t_i)\pi^2 \hbar^{\frac{3}{2}}}{\sqrt{2N(t_f-t_i)}} =1 ,
\end{align}
and the normalized transition amplitude \eqref{eq:action-jt-minisuperspace_re} is
\begin{align}
    P(N;\; \Pi_{q}^{1}, \Pi_{\varphi}^{1}; \;\Pi_{q}^{0}, \Pi_{\varphi}^{0}) 
    =\exp\Bigg[\frac{2iN(t_f-t_i)}{\pi\hbar} \Pi_{\varphi}^{1} \Pi_{q}^{1}
    -\frac{2i\pi}{\hbar} N \Lambda (t_f-t_i)
    \Bigg]
    \delta \Big[\Pi_{\varphi}^{1} -\Pi_{\varphi}^{0} \Big]
    \delta \Big[\Pi_{q}^{1} -\Pi_{q}^{0} \Big]
    \label{eq:TA_NN_nom}
\end{align}
Remarkably, this amplitude gives a non-zero value for only trivial transition with $\Pi_{\varphi}^{1} = \Pi_{\varphi}^{0}$ and $\Pi_{q}^{1} =\Pi_{q}^{0}$. This is consistent with the current system without the potential of $q$ or $\varphi$. 

\subsection{Neumann-Dirichlet and Dirichlet-Neumann}
Here we consider the case with initially Neumann- and finally Dirichlet-type boundary conditions:
\begin{equation}
    -\frac{\pi}{2\tilde{N}}\dot{q}(0)=\Pi_{\varphi}^{0}, \quad
    -\frac{\pi}{2\tilde{N}}\dot{\varphi}(0)=\Pi_{q}^{0}, \quad
    q(1)=q_{1}, \quad
    \varphi(1)=\varphi_{1}.
\end{equation} 
The corresponding boundary action is given as
\begin{align}
    S_B
    :=
    \Pi_{\varphi}^{0} \varphi(0)+q(0) \Pi_{q}^{0}
    \,.
\end{align}
Note that the boundary term for the Dirichlet one is unnecessary. In this case, the following redefinition of the fields is useful:
\begin{align}
    \tilde{q}(t)
    := q(t) + \frac{2\tilde{N}}{\pi} \Pi_{\varphi}^{0} \, (t-1) - q_{1}, \quad
    \tilde{\varphi}(t)
    := \varphi(t) + \frac{2\tilde{N}}{\pi} \Pi_{q}^{0} \, (t-1) - \varphi_{1},
\end{align}
which satisfies
\begin{align}
    \dot{\tilde{q}}(0) = \dot{\tilde{\varphi}}(0) = 0, \quad
    \tilde{q}(1) = \tilde{\varphi}(1) = 0.
\end{align}
Then we can perform the Fourier decomposition of these redefined fields
\begin{align}
    \tilde{q}(t) =\sum_{n=0}^\infty \tilde{q}_n \cos\Bigg[\left(n+\frac{1}{2}\right)\pi t\Bigg] ,\quad
    \tilde{\varphi}(t)= \sum_{n=0}^\infty \tilde{\varphi}_n \cos\Bigg[\left(n+\frac{1}{2}\right)\pi t\Bigg]  .
\end{align}
The bulk action is reduced to
\begin{align}
    S_{q\varphi}
    &=
    -\frac{\pi^3}{4 \tilde{N}} \sum_{n=0}^{\infty} 
    \tilde{q}_n \tilde{\varphi}_n \left(n+\frac{1}{2}\right)^2
    - \frac{2\tilde{N}}{\pi} \Pi_{q}^{0} \Pi_{\varphi}^{0}
    -\Big(
    \Pi_{\varphi}^{0} \tilde{\varphi}(0)+\tilde{q}(0) \Pi_{q}^{0}
    \Big),
\end{align}
and the entire action is obtained by adding boundary action
\begin{align}
    S_{q\varphi} + S_{B}
    &=
    -\frac{\pi^3}{4 \tilde{N}} \sum_{n=0}^{\infty} 
    \tilde{q}_n \tilde{\varphi}_n \left(n+\frac{1}{2}\right)^2
    +\frac{2\tilde{N}}{\pi} \Pi_{q}^{0} \Pi_{\varphi}^{0}
    + \Pi_{\varphi}^{0} \varphi_1 + \Pi_{q}^{0} q_1.
\end{align}

Then the transition amplitude is
\begin{align}
    P_{q\varphi}
    (N;\; q_1, \varphi_1; \;\Pi_{q}^{0}, \Pi_{\varphi}^{0}) 
    &= C_{ND} (t_f,t_i) \exp
    \Bigg[
    \frac{i}{\hbar} \left(
    \frac{2\tilde{N}}{\pi} \Pi_{q}^{0} \Pi_{\varphi}^{0}
    + \Pi_{\varphi}^{0} \varphi_1 + \Pi_{q}^{0} q_1
    \right)
    \Bigg]
    \nonumber\\
    & \hspace{120pt} 
    \int  \left( \prod_{n=0}^\infty \mathrm{d}\tilde{q}_n~\mathrm{d}\tilde{\varphi}_n  \right) \exp\Bigg[ 
        -\frac{i\pi^3}{4 \hbar \tilde{N}} \sum_{n=0}^{\infty} 
        \tilde{q}_n \tilde{\varphi}_n \left(n+\frac{1}{2}\right)^2
    \Bigg] 
    \nonumber \\ 
    &= \frac{\pi C_{ND} (t_f,t_i)}{\sqrt{2\hbar N(t_f-t_i)}} 
    \exp \Bigg[
    \frac{i}{\hbar} \left(
    \frac{2N(t_f-t_i)}{\pi} \Pi_{q}^{0} \Pi_{\varphi}^{0}
    + \Pi_{\varphi}^{0} \varphi_1 + \Pi_{q}^{0} q_1
    \right)
    \Bigg],
\end{align}
where we performed zeta-function regularization \eqref{eq:zeta_reg} and
\begin{align}
    \prod_{n=0}^{\infty} \frac{1}{\left(n+\frac{1}{2}\right)^2}
    =\exp \left[-2 \sum_{n=0}^{\infty} \log \left(n+\frac{1}{2}\right)\right]
    =\exp [\log (2)]=2.
\end{align}

The normalization $C_{ND}(t_f,t_i)$ cannot be determined alone. Hence we normalize by combining the amplitude under the Dirichlet-Neumann condition
\begin{equation}
    -\frac{\pi}{2\tilde{N}}\dot{q}(1)=\Pi_{\varphi}^{1}, \quad
    -\frac{\pi}{2\tilde{N}}\dot{\varphi}(1)=\Pi_{q}^{1}, \quad
    q(0)=q_{0}, \quad
    \varphi(0)=\varphi_{0}.
\end{equation} 
Similar calculation leads to the corresponding amplitude
\begin{align}
    P_{q\varphi}(N;\; \Pi_{q}^{1}, \Pi_{\varphi}^{1}\; q_0, \varphi_0) 
    &= \frac{\pi C_{ DN}(t_f,t_i)}{\sqrt{2\hbar N(t_f-t_i)}} 
    \exp \Bigg[
    \frac{i}{\hbar} \left(
    \frac{2N(t_f-t_i)}{\pi} \Pi_{q}^{1} \Pi_{\varphi}^{1}
    - \Pi_{\varphi}^{1} \varphi_0 - \Pi_{q}^{1} q_0
    \right)
    \Bigg],
\end{align}
Then we set the normalization condition as
\begin{align}
    &\int \mathrm{d}p_q \mathrm{d}p_\varphi \;
    P_{q\varphi}(N;\; q_1, \varphi_1\; p_{q}, p_{\varphi}) 
    P_{q\varphi}(N;\; p_{q}, p_{\varphi}\; q_0, \varphi_0) 
    = P_{q\varphi}(N;\; q_1, \varphi_1;\; q_0, \varphi_0) , \\
    &
    \int \mathrm{d}q \mathrm{d}\varphi \; 
    P_{q\varphi}(N;\; \Pi_{q}^{1}, \Pi_{\varphi}^{1}\; q, \varphi) 
    P(_{q\varphi}N;\; q, \varphi \; \Pi_{q}^{1}, \Pi_{\varphi}^{1}) 
    = P_{q\varphi}(N;\; \Pi_{q}^{1}, \Pi_{\varphi}^{1}; \; \Pi_{q}^{0}, \Pi_{\varphi}^{0}) .
\end{align}
From the first condition
\begin{align}
    (\mathrm{L.H.S})
    &= \frac{\pi^2 C_{ND}(t_f,t) C_{DN}(t,t_i)}{2\hbar N\sqrt{(t_f-t)(t-t_i)}}
    \int \mathrm{d}p_q \mathrm{d}p_\varphi \;
    \exp \Bigg[
    \frac{i}{\hbar} \left(
    \frac{2N(t_f-t)}{\pi} p_{q} p_{\varphi}
    + p_{\varphi} \varphi_1 + p_{q} q_1
    \right)
    \Bigg]
    \nonumber\\
    &\hspace{150pt}
    \exp \Bigg[
    \frac{i}{\hbar} \left(
    \frac{2N(t-t_i)}{\pi} p_{q} p_{\varphi}
    - p_{\varphi} \varphi_0 - p_{q} q_0
    \right)
    \Bigg]
    \nonumber\\
    &= \frac{\pi^4 C_{ND}(t_f,t) C_{DN}(t,t_i)}{2N^2(t_f-t_i)\sqrt{(t_f-t)(t-t_i)}}
    \exp \Bigg[
    -\frac{i\pi}{2\hbar N(t_f-t_i)} 
    (q_1 -q_0) (\varphi_1 -\varphi_0)
    \Bigg]
    \nonumber\\
    (\mathrm{R.H.S})
    &= \frac{1}{ 4\hbar N(t_f-t_i)} 
    \exp \Bigg[
    -\frac{i\pi}{2\hbar N(t_f-t_i)} 
    (q_1 -q_0) (\varphi_1 -\varphi_0)
    \Bigg].
    \label{eq:DNND=DD}
\end{align}
From the second condition
\begin{align}
    (\mathrm{L.H.S})
    &= \frac{\pi^2 C_{DN}(t_f,t) C_{
    ND}(t,t_i)}{2\hbar N\sqrt{(t_f-t)(t-t_i)}}
    \int \mathrm{d}q \mathrm{d}\varphi \;
    \exp \Bigg[
    \frac{i}{\hbar} \left(
    \frac{2N(t-t_i)}{\pi} \Pi_{q}^{0} \Pi_{\varphi}^{0}
    + \Pi_{\varphi}^{0} \varphi + \Pi_{q}^{0} q
    \right)
    \Bigg]
    \nonumber\\
    &\hspace{150pt}
    \cdot\exp \Bigg[
    \frac{i}{\hbar} \left(
    \frac{2N(t_f-t)}{\pi} \Pi_{q}^{1} \Pi_{\varphi}^{1}
    - \Pi_{\varphi}^{1} \varphi - \Pi_{q}^{1} q
    \right)
    \Bigg]
    \nonumber\\
    &=\frac{2\pi^4 \hbar \, C_{DN}(t_f,t) C_{ND}(t,t_i)}{N\sqrt{(t_f-t)(t-t_i)}}
    \exp \Bigg[
    \frac{2iN(t_f-t_i)}{\pi\hbar} \Pi_{q}^{0} \Pi_{\varphi}^{0}
    \Bigg]
    \delta \Big[\Pi_{\varphi}^{1} -\Pi_{\varphi}^{0} \Big]
    \delta \Big[\Pi_{q}^{1} -\Pi_{q}^{0} \Big],
    \nonumber\\
    (\mathrm{R.H.S})
    &= \exp\Bigg[\frac{2iN(t_f-t_i)}{\pi\hbar} \Pi_{\varphi}^{1} \Pi_{q}^{1}\Bigg]
    \delta \Big[\Pi_{\varphi}^{1} -\Pi_{\varphi}^{0} \Big]
    \delta \Big[\Pi_{q}^{1} -\Pi_{q}^{0} \Big].
    \label{eq:NDDN=NN}
\end{align}
Here we used the results of the Dirichlet-Dirichlet case \eqref{eq:TA_DD_nom} and the Neumann-Neumann case \eqref{eq:TA_NN_nom}. These conditions suggest that
\begin{align}
    C_{ND} C_{DN} = \frac{N\sqrt{(t_f-t)(t-t_i)}}{2\pi^4 \hbar}.
\end{align}
If we take $C_{ND}$ and $C_{DN}$ having the same value, 
\begin{align}
    C_{ND} (t_f, t_i)
    =C_{DN} (t_f, t_i)
    = \sqrt{\frac{N(t_f-t_i)}{2\pi^4 \hbar}}
\end{align}
the whole normalized amplitudes \eqref{eq:action-jt-minisuperspace_re} result in
\begin{align}
    P(N;\; q_1, \varphi_1; \;\Pi_{q}^{0}, \Pi_{\varphi}^{0}) 
    &= \frac{1}{2\pi \hbar} 
    \exp \Bigg[
    \frac{i}{\hbar} \left(
    \frac{2N(t_f-t_i)}{\pi} \Pi_{q}^{0} \Pi_{\varphi}^{0}
    + \Pi_{\varphi}^{0} \varphi_1 + \Pi_{q}^{0} q_1
    -2\pi N \Lambda (t_f-t_i)
    \right)
    \Bigg], \\
    P(N;\; \Pi_{q}^{1}, \Pi_{\varphi}^{1}\; q_0, \varphi_0) 
    &= \frac{1}{2\pi \hbar} 
    \exp \Bigg[
    \frac{i}{\hbar} \left(
    \frac{2N(t_f-t_i)}{\pi} \Pi_{q}^{1} \Pi_{\varphi}^{1}
    - \Pi_{\varphi}^{1} \varphi_0 - \Pi_{q}^{1} q_0
    -2\pi N \Lambda (t_f-t_i)
    \right)
    \Bigg],
\end{align}
respectively. The lapse $N$ is absent in the consequent amplitudes of the Dirichlet-Neumann and Neumann-Dirichlet case. This is consistent with the 4-D GR result \cite{Honda:2024aro}.

\twocolumngrid

\bibliographystyle{utphys}
\bibliography{Refs.bib}

\providecommand{\href}[2]{#2}\begingroup\raggedright\begin{thebibliography}{10}

\bibitem{Jackiw:1984je}
R.~Jackiw, ``{Lower Dimensional Gravity},''
  \href{http://dx.doi.org/10.1016/0550-3213(85)90448-1}{{\em Nucl. Phys. B}
  {\bfseries 252} (1985) 343--356}.

\bibitem{Teitelboim:1983ux}
C.~Teitelboim, ``{Gravitation and Hamiltonian Structure in Two Space-Time
  Dimensions},'' \href{http://dx.doi.org/10.1016/0370-2693(83)90012-6}{{\em
  Phys. Lett. B} {\bfseries 126} (1983) 41--45}.

\bibitem{Almheiri:2014cka}
A.~Almheiri and J.~Polchinski, ``{Models of AdS$_{2}$ backreaction and
  holography},'' \href{http://dx.doi.org/10.1007/JHEP11(2015)014}{{\em JHEP}
  {\bfseries 11} (2015) 014}, \href{http://arxiv.org/abs/1402.6334}{{\ttfamily
  arXiv:1402.6334 [hep-th]}}.

\bibitem{Jensen:2016pah}
K.~Jensen, ``{Chaos in AdS$_2$ Holography},''
  \href{http://dx.doi.org/10.1103/PhysRevLett.117.111601}{{\em Phys. Rev.
  Lett.} {\bfseries 117} no.~11, (2016) 111601},
  \href{http://arxiv.org/abs/1605.06098}{{\ttfamily arXiv:1605.06098
  [hep-th]}}.

\bibitem{Maldacena:2016upp}
J.~Maldacena, D.~Stanford, and Z.~Yang, ``{Conformal symmetry and its breaking
  in two dimensional Nearly Anti-de-Sitter space},''
  \href{http://dx.doi.org/10.1093/ptep/ptw124}{{\em PTEP} {\bfseries 2016}
  no.~12, (2016) 12C104}, \href{http://arxiv.org/abs/1606.01857}{{\ttfamily
  arXiv:1606.01857 [hep-th]}}.

\bibitem{Engelsoy:2016xyb}
J.~Engels\"oy, T.~G. Mertens, and H.~Verlinde, ``{An investigation of AdS$_{2}$
  backreaction and holography},''
  \href{http://dx.doi.org/10.1007/JHEP07(2016)139}{{\em JHEP} {\bfseries 07}
  (2016) 139}, \href{http://arxiv.org/abs/1606.03438}{{\ttfamily
  arXiv:1606.03438 [hep-th]}}.

\bibitem{Nayak:2018qej}
P.~Nayak, A.~Shukla, R.~M. Soni, S.~P. Trivedi, and V.~Vishal, ``{On the
  Dynamics of Near-Extremal Black Holes},''
  \href{http://dx.doi.org/10.1007/JHEP09(2018)048}{{\em JHEP} {\bfseries 09}
  (2018) 048}, \href{http://arxiv.org/abs/1802.09547}{{\ttfamily
  arXiv:1802.09547 [hep-th]}}.

\bibitem{Grumiller:2016dbn}
D.~Grumiller, J.~Salzer, and D.~Vassilevich, ``{Aspects of AdS$_2$ holography
  with non-constant dilaton},''
  \href{http://dx.doi.org/10.1007/s11182-017-0978-x}{{\em Russ. Phys. J.}
  {\bfseries 59} no.~11, (2017) 1798--1803},
  \href{http://arxiv.org/abs/1607.06974}{{\ttfamily arXiv:1607.06974
  [hep-th]}}.

\bibitem{Mertens:2022irh}
T.~G. Mertens and G.~J. Turiaci, ``{Solvable models of quantum black holes: a
  review on Jackiw\textendash{}Teitelboim gravity},''
  \href{http://dx.doi.org/10.1007/s41114-023-00046-1}{{\em Living Rev. Rel.}
  {\bfseries 26} no.~1, (2023) 4},
  \href{http://arxiv.org/abs/2210.10846}{{\ttfamily arXiv:2210.10846
  [hep-th]}}.

\bibitem{Henneaux:1985nw}
M.~Henneaux, ``{QUANTUM GRAVITY IN TWO-DIMENSIONS: EXACT SOLUTION OF THE JACKIW
  MODEL},'' \href{http://dx.doi.org/10.1103/PhysRevLett.54.959}{{\em Phys. Rev.
  Lett.} {\bfseries 54} (1985) 959--962}.

\bibitem{LouisMartinez:1993eh}
D.~Louis-Martinez, J.~Gegenberg, and G.~Kunstatter, ``{Exact Dirac quantization
  of all 2-D dilaton gravity theories},''
  \href{http://dx.doi.org/10.1016/0370-2693(94)90463-4}{{\em Phys. Lett. B}
  {\bfseries 321} (1994) 193--198},
  \href{http://arxiv.org/abs/gr-qc/9309018}{{\ttfamily arXiv:gr-qc/9309018}}.

\bibitem{Maldacena:2019cbz}
J.~Maldacena, G.~J. Turiaci, and Z.~Yang, ``{Two dimensional Nearly de Sitter
  gravity},'' \href{http://dx.doi.org/10.1007/JHEP01(2021)139}{{\em JHEP}
  {\bfseries 01} (2021) 139}, \href{http://arxiv.org/abs/1904.01911}{{\ttfamily
  arXiv:1904.01911 [hep-th]}}.

\bibitem{Stanford:2019vob}
D.~Stanford and E.~Witten, ``{JT gravity and the ensembles of random matrix
  theory},'' \href{http://dx.doi.org/10.4310/ATMP.2020.v24.n6.a4}{{\em Adv.
  Theor. Math. Phys.} {\bfseries 24} no.~6, (2020) 1475--1680},
  \href{http://arxiv.org/abs/1907.03363}{{\ttfamily arXiv:1907.03363
  [hep-th]}}.

\bibitem{Cotler:2019nbi}
J.~Cotler, K.~Jensen, and A.~Maloney, ``{Low-dimensional de Sitter quantum
  gravity},'' \href{http://dx.doi.org/10.1007/JHEP06(2020)048}{{\em JHEP}
  {\bfseries 06} (2020) 048}, \href{http://arxiv.org/abs/1905.03780}{{\ttfamily
  arXiv:1905.03780 [hep-th]}}.

\bibitem{Iliesiu:2020zld}
L.~V. Iliesiu, J.~Kruthoff, G.~J. Turiaci, and H.~Verlinde, ``{JT gravity at
  finite cutoff},'' \href{http://dx.doi.org/10.21468/SciPostPhys.9.2.023}{{\em
  SciPost Phys.} {\bfseries 9} (2020) 023},
  \href{http://arxiv.org/abs/2004.07242}{{\ttfamily arXiv:2004.07242
  [hep-th]}}.

\bibitem{Stanford:2020qhm}
D.~Stanford and Z.~Yang, ``{Finite-cutoff JT gravity and self-avoiding
  loops},'' \href{http://arxiv.org/abs/2004.08005}{{\ttfamily arXiv:2004.08005
  [hep-th]}}.

\bibitem{Moitra:2021uiv}
U.~Moitra, S.~K. Sake, and S.~P. Trivedi, ``{Jackiw-Teitelboim gravity in the
  second order formalism},''
  \href{http://dx.doi.org/10.1007/JHEP10(2021)204}{{\em JHEP} {\bfseries 10}
  (2021) 204}, \href{http://arxiv.org/abs/2101.00596}{{\ttfamily
  arXiv:2101.00596 [hep-th]}}.

\bibitem{Anegawa:2023wrk}
T.~Anegawa, N.~Iizuka, S.~K. Sake, and N.~Zenoni, ``{Is action complexity
  better for de Sitter space in Jackiw-Teitelboim gravity?},''
  \href{http://dx.doi.org/10.1007/JHEP06(2023)213}{{\em JHEP} {\bfseries 06}
  (2023) 213}, \href{http://arxiv.org/abs/2303.05025}{{\ttfamily
  arXiv:2303.05025 [hep-th]}}.

\bibitem{Nanda:2023wne}
K.~K. Nanda, S.~K. Sake, and S.~P. Trivedi, ``{JT gravity in de Sitter space
  and the problem of time},''
  \href{http://dx.doi.org/10.1007/JHEP02(2024)145}{{\em JHEP} {\bfseries 02}
  (2024) 145}, \href{http://arxiv.org/abs/2307.15900}{{\ttfamily
  arXiv:2307.15900 [hep-th]}}.

\bibitem{Buchmuller:2024ksd}
W.~Buchmuller, A.~Hebecker, and A.~Westphal, ``{DeWitt wave functions for de
  Sitter JT gravity},'' \href{http://arxiv.org/abs/2412.09211}{{\ttfamily
  arXiv:2412.09211 [hep-th]}}.

\bibitem{Hartle:1983ai}
J.~B. Hartle and S.~W. Hawking, ``{Wave Function of the Universe},''
  \href{http://dx.doi.org/10.1103/PhysRevD.28.2960}{{\em Phys. Rev. D}
  {\bfseries 28} (1983) 2960--2975}.

\bibitem{Vilenkin:1982de}
A.~Vilenkin, ``{Creation of Universes from Nothing},''
  \href{http://dx.doi.org/10.1016/0370-2693(82)90866-8}{{\em Phys. Lett. B}
  {\bfseries 117} (1982) 25--28}.

\bibitem{Gibbons:1978ac}
G.~W. Gibbons, S.~W. Hawking, and M.~J. Perry, ``{Path Integrals and the
  Indefiniteness of the Gravitational Action},''
  \href{http://dx.doi.org/10.1016/0550-3213(78)90161-X}{{\em Nucl. Phys. B}
  {\bfseries 138} (1978) 141--150}.

\bibitem{Linde:1983mx}
A.~D. Linde, ``{Quantum Creation of the Inflationary Universe},''
  \href{http://dx.doi.org/10.1007/BF02790571}{{\em Lett. Nuovo Cim.} {\bfseries
  39} (1984) 401--405}.

\bibitem{Halliwell:1988ik}
J.~J. Halliwell and J.~Louko, ``{Steepest Descent Contours in the Path Integral
  Approach to Quantum Cosmology. 1. The De Sitter Minisuperspace Model},''
  \href{http://dx.doi.org/10.1103/PhysRevD.39.2206}{{\em Phys. Rev. D}
  {\bfseries 39} (1989) 2206}.

\bibitem{Feldbrugge:2017kzv}
J.~Feldbrugge, J.-L. Lehners, and N.~Turok, ``{Lorentzian Quantum Cosmology},''
  \href{http://dx.doi.org/10.1103/PhysRevD.95.103508}{{\em Phys. Rev. D}
  {\bfseries 95} no.~10, (2017) 103508},
  \href{http://arxiv.org/abs/1703.02076}{{\ttfamily arXiv:1703.02076
  [hep-th]}}.

\bibitem{Pham}
F.~PHAM, ``Vanishing homologies and the n variables saddlepoint method,'' {\em
  Proc. Symp. Pure Math.} {\bfseries 40} (1983) 310--333.

\bibitem{Berry:1991}
M.~V. Berry and C.~J. Howls, ``Hyperasymptotics for integrals with saddles,''
  {\em Proceedings: Mathematical and Physical Sciences} {\bfseries 434}
  no.~1892, (1991) 657--675.

\bibitem{Howls}
C.~J. Howls, ``Hyperasymptotics for multidimensional integrals, exact remainder
  terms and the global connection problem,''
  \href{http://dx.doi.org/10.1098/rspa.1997.0122}{{\em Proceedings of the Royal
  Society of London. Series A: Mathematical, Physical and Engineering Sciences}
  {\bfseries 453} no.~1966, (1997) 2271--2294}.

\bibitem{Witten:2010cx}
E.~Witten, ``{Analytic Continuation Of Chern-Simons Theory},'' {\em AMS/IP
  Stud. Adv. Math.} {\bfseries 50} (2011) 347--446,
  \href{http://arxiv.org/abs/1001.2933}{{\ttfamily arXiv:1001.2933 [hep-th]}}.

\bibitem{Mou:2019tck}
Z.-G. Mou, P.~M. Saffin, A.~Tranberg, and S.~Woodward, ``{Real-time quantum
  dynamics, path integrals and the method of thimbles},''
  \href{http://dx.doi.org/10.1007/JHEP06(2019)094}{{\em JHEP} {\bfseries 06}
  (2019) 094}, \href{http://arxiv.org/abs/1902.09147}{{\ttfamily
  arXiv:1902.09147 [hep-lat]}}.

\bibitem{Mou:2019gyl}
Z.-G. Mou, P.~M. Saffin, and A.~Tranberg, ``{Quantum tunnelling, real-time
  dynamics and Picard-Lefschetz thimbles},''
  \href{http://dx.doi.org/10.1007/JHEP11(2019)135}{{\em JHEP} {\bfseries 11}
  (2019) 135}, \href{http://arxiv.org/abs/1909.02488}{{\ttfamily
  arXiv:1909.02488 [hep-th]}}.

\bibitem{Millington:2020vkg}
P.~Millington, Z.-G. Mou, P.~M. Saffin, and A.~Tranberg, ``{Statistics on
  Lefschetz thimbles: Bell/Leggett-Garg inequalities and the
  classical-statistical approximation},''
  \href{http://dx.doi.org/10.1007/JHEP03(2021)077}{{\em JHEP} {\bfseries 03}
  (2021) 077}, \href{http://arxiv.org/abs/2011.02657}{{\ttfamily
  arXiv:2011.02657 [hep-th]}}.

\bibitem{Matsui:2021oio}
H.~Matsui, ``{Lorentzian path integral for quantum tunneling and WKB
  approximation for wave-function},''
  \href{http://dx.doi.org/10.1140/epjc/s10052-022-10374-1}{{\em Eur. Phys. J.
  C} {\bfseries 82} no.~5, (2022) 426},
  \href{http://arxiv.org/abs/2102.09767}{{\ttfamily arXiv:2102.09767 [gr-qc]}}.

\bibitem{Rajeev:2021zae}
K.~Rajeev, ``{Lorentzian worldline path integral approach to Schwinger
  effect},'' \href{http://dx.doi.org/10.1103/PhysRevD.104.105014}{{\em Phys.
  Rev. D} {\bfseries 104} no.~10, (2021) 105014},
  \href{http://arxiv.org/abs/2105.12194}{{\ttfamily arXiv:2105.12194
  [hep-th]}}.

\bibitem{Hayashi:2021kro}
T.~Hayashi, K.~Kamada, N.~Oshita, and J.~Yokoyama, ``{Vacuum decay in the
  Lorentzian path integral},''
  \href{http://dx.doi.org/10.1088/1475-7516/2022/05/041}{{\em JCAP} {\bfseries
  05} no.~05, (2022) 041}, \href{http://arxiv.org/abs/2112.09284}{{\ttfamily
  arXiv:2112.09284 [hep-th]}}.

\bibitem{Feldbrugge:2022idb}
J.~Feldbrugge and N.~Turok, ``{Existence of real time quantum path
  integrals},'' \href{http://dx.doi.org/10.1016/j.aop.2023.169315}{{\em Annals
  Phys.} {\bfseries 454} (2023) 169315},
  \href{http://arxiv.org/abs/2207.12798}{{\ttfamily arXiv:2207.12798
  [hep-th]}}.

\bibitem{Nishimura:2023dky}
J.~Nishimura, K.~Sakai, and A.~Yosprakob, ``{A new picture of quantum tunneling
  in the real-time path integral from Lefschetz thimble calculations},''
  \href{http://dx.doi.org/10.1007/JHEP09(2023)110}{{\em JHEP} {\bfseries 09}
  (2023) 110}, \href{http://arxiv.org/abs/2307.11199}{{\ttfamily
  arXiv:2307.11199 [hep-th]}}.

\bibitem{Feldbrugge:2023frq}
J.~Feldbrugge, D.~L. Jow, and U.-L. Pen, ``{Complex classical paths in quantum
  reflections and tunneling},''
  \href{http://arxiv.org/abs/2309.12420}{{\ttfamily arXiv:2309.12420
  [quant-ph]}}.

\bibitem{Feldbrugge:2023mhn}
J.~Feldbrugge, D.~L. Jow, and U.-L. Pen, ``{Crossing singularities in the
  saddle point approximation},''
  \href{http://arxiv.org/abs/2309.12427}{{\ttfamily arXiv:2309.12427
  [quant-ph]}}.

\bibitem{Saito:2024acm}
D.~Saito and N.~Oshita, ``{Remote Hawking-Moss instanton and the Lorentzian
  path integral},'' \href{http://arxiv.org/abs/2409.03978}{{\ttfamily
  arXiv:2409.03978 [hep-th]}}.

\bibitem{Lehners:2023yrj}
J.-L. Lehners, ``{Review of the no-boundary wave function},''
  \href{http://dx.doi.org/10.1016/j.physrep.2023.06.002}{{\em Phys. Rept.}
  {\bfseries 1022} (2023) 1--82},
  \href{http://arxiv.org/abs/2303.08802}{{\ttfamily arXiv:2303.08802
  [hep-th]}}.

\bibitem{Lehners:2024kus}
J.-L. Lehners, ``{NUTs, Bolts and Stokes Phenomena in the No-Boundary Wave
  Function},'' \href{http://arxiv.org/abs/2402.10501}{{\ttfamily
  arXiv:2402.10501 [gr-qc]}}.

\bibitem{Honda:2024aro}
M.~Honda, H.~Matsui, K.~Okabayashi, and T.~Terada, ``{Resurgence in Lorentzian
  quantum cosmology: No-boundary saddles and resummation of quantum gravity
  corrections around tunneling saddle points},''
  \href{http://dx.doi.org/10.1103/PhysRevD.110.083508}{{\em Phys. Rev. D}
  {\bfseries 110} no.~8, (2024) 083508},
  \href{http://arxiv.org/abs/2402.09981}{{\ttfamily arXiv:2402.09981 [gr-qc]}}.

\bibitem{Chou:2024sgk}
C.-Y. Chou and J.~Nishimura, ``{Monte Carlo studies of quantum cosmology by the
  generalized Lefschetz thimble method},''
  \href{http://arxiv.org/abs/2407.17724}{{\ttfamily arXiv:2407.17724 [gr-qc]}}.

\bibitem{Matsui:2023hei}
H.~Matsui and S.~Mukohyama, ``{Hartle-Hawking no-boundary proposal and
  Ho\v{r}ava-Lifshitz gravity},''
  \href{http://dx.doi.org/10.1103/PhysRevD.109.023504}{{\em Phys. Rev. D}
  {\bfseries 109} no.~2, (2024) 023504},
  \href{http://arxiv.org/abs/2310.00210}{{\ttfamily arXiv:2310.00210 [gr-qc]}}.

\bibitem{chen_hikida_taki:2024short}
H.-Y. Chen, Y.~Hikida, Y.~Taki, and T.~Uetoko, ``Semi-classical saddles of
  three-dimensional gravity via holography,'' 2024.
\newblock \url{https://arxiv.org/abs/2403.02108}.

\bibitem{chen_hikida_taki:2024long}
H.-Y. Chen, Y.~Hikida, Y.~Taki, and T.~Uetoko, ``The semi-classical saddles in
  three-dimensional gravity via holography and mini-superspace approach,''
  2024.
\newblock \url{https://arxiv.org/abs/2404.10277}.

\bibitem{Halliwell:1990tu}
J.~J. Halliwell and J.~Louko, ``{Steepest Descent Contours in the Path Integral
  Approach to Quantum Cosmology. 3. A General Method With Applications to
  Anisotropic Minisuperspace Models},''
  \href{http://dx.doi.org/10.1103/PhysRevD.42.3997}{{\em Phys. Rev. D}
  {\bfseries 42} (1990) 3997--4031}.

\bibitem{Fanaras:2021awm}
G.~Fanaras and A.~Vilenkin, ``{Jackiw-Teitelboim and Kantowski-Sachs quantum
  cosmology},'' \href{http://dx.doi.org/10.1088/1475-7516/2022/03/056}{{\em
  JCAP} {\bfseries 03} no.~03, (2022) 056},
  \href{http://arxiv.org/abs/2112.00919}{{\ttfamily arXiv:2112.00919 [gr-qc]}}.

\bibitem{Fanaras:2022twv}
G.~Fanaras and A.~Vilenkin, ``{The tunneling wavefunction in Kantowski-Sachs
  quantum cosmology},''
  \href{http://dx.doi.org/10.1088/1475-7516/2022/08/069}{{\em JCAP} {\bfseries
  08} (2022) 069}, \href{http://arxiv.org/abs/2206.05839}{{\ttfamily
  arXiv:2206.05839 [gr-qc]}}.

\bibitem{Ghosh:2023njl}
S.~Ghosh, A.~Acharya, S.~Gangopadhyay, and P.~K. Panigrahi, ``{Lorentzian path
  integral in Kantowski-Sachs anisotropic cosmology},''
  \href{http://dx.doi.org/10.1103/PhysRevD.109.043524}{{\em Phys. Rev. D}
  {\bfseries 109} no.~4, (2024) 043524},
  \href{http://arxiv.org/abs/2310.15180}{{\ttfamily arXiv:2310.15180 [gr-qc]}}.

\bibitem{Louko:1988bk}
J.~Louko, ``{Canonizing the Hartle-hawking Proposal},''
  \href{http://dx.doi.org/10.1016/0370-2693(88)90008-1}{{\em Phys. Lett. B}
  {\bfseries 202} (1988) 201--206}.

\bibitem{Conradi:1994yy}
H.-D. Conradi, ``{Quantum cosmology of Kantowski-Sachs like models},''
  \href{http://dx.doi.org/10.1088/0264-9381/12/10/005}{{\em Class. Quant.
  Grav.} {\bfseries 12} (1995) 2423--2440},
  \href{http://arxiv.org/abs/gr-qc/9412049}{{\ttfamily arXiv:gr-qc/9412049}}.

\bibitem{Louko:1988ia}
J.~Louko and T.~Vachaspati, ``{On the Vilenkin Boundary Condition Proposal in
  Anisotropic Universes},''
  \href{http://dx.doi.org/10.1016/0370-2693(89)90912-X}{{\em Phys. Lett. B}
  {\bfseries 223} (1989) 21--25}.

\bibitem{Conti:2014uda}
G.~Conti and T.~Hertog, ``{Two wave functions and dS/CFT on S$^{1}$
  \texttimes{} S$^{2}$},''
  \href{http://dx.doi.org/10.1007/JHEP06(2015)101}{{\em JHEP} {\bfseries 06}
  (2015) 101}, \href{http://arxiv.org/abs/1412.3728}{{\ttfamily arXiv:1412.3728
  [hep-th]}}.

\bibitem{Stanford:2017thb}
D.~Stanford and E.~Witten, ``{Fermionic Localization of the Schwarzian
  Theory},'' \href{http://dx.doi.org/10.1007/JHEP10(2017)008}{{\em JHEP}
  {\bfseries 10} (2017) 008}, \href{http://arxiv.org/abs/1703.04612}{{\ttfamily
  arXiv:1703.04612 [hep-th]}}.

\bibitem{Halliwell:1988wc}
J.~J. Halliwell, ``{Derivation of the Wheeler-De Witt Equation from a Path
  Integral for Minisuperspace Models},''
  \href{http://dx.doi.org/10.1103/PhysRevD.38.2468}{{\em Phys. Rev. D}
  {\bfseries 38} (1988) 2468}.

\bibitem{Teitelboim:1983fh}
C.~Teitelboim, ``{Causality Versus Gauge Invariance in Quantum Gravity and
  Supergravity},'' \href{http://dx.doi.org/10.1103/PhysRevLett.50.705}{{\em
  Phys. Rev. Lett.} {\bfseries 50} (1983) 705}.

\bibitem{DiazDorronsoro:2017hti}
J.~Diaz~Dorronsoro, J.~J. Halliwell, J.~B. Hartle, T.~Hertog, and O.~Janssen,
  ``{Real no-boundary wave function in Lorentzian quantum cosmology},''
  \href{http://dx.doi.org/10.1103/PhysRevD.96.043505}{{\em Phys. Rev. D}
  {\bfseries 96} no.~4, (2017) 043505},
  \href{http://arxiv.org/abs/1705.05340}{{\ttfamily arXiv:1705.05340 [gr-qc]}}.

\bibitem{Feldbrugge:2017mbc}
J.~Feldbrugge, J.-L. Lehners, and N.~Turok, ``{No rescue for the no boundary
  proposal: Pointers to the future of quantum cosmology},''
  \href{http://dx.doi.org/10.1103/PhysRevD.97.023509}{{\em Phys. Rev. D}
  {\bfseries 97} no.~2, (2018) 023509},
  \href{http://arxiv.org/abs/1708.05104}{{\ttfamily arXiv:1708.05104
  [hep-th]}}.

\bibitem{DiTucci:2019dji}
A.~Di~Tucci and J.-L. Lehners, ``{No-Boundary Proposal as a Path Integral with
  Robin Boundary Conditions},''
  \href{http://dx.doi.org/10.1103/PhysRevLett.122.201302}{{\em Phys. Rev.
  Lett.} {\bfseries 122} no.~20, (2019) 201302},
  \href{http://arxiv.org/abs/1903.06757}{{\ttfamily arXiv:1903.06757
  [hep-th]}}.

\bibitem{DiTucci:2019bui}
A.~Di~Tucci, J.-L. Lehners, and L.~Sberna, ``{No-boundary prescriptions in
  Lorentzian quantum cosmology},''
  \href{http://dx.doi.org/10.1103/PhysRevD.100.123543}{{\em Phys. Rev. D}
  {\bfseries 100} no.~12, (2019) 123543},
  \href{http://arxiv.org/abs/1911.06701}{{\ttfamily arXiv:1911.06701
  [hep-th]}}.

\bibitem{Narain:2021bff}
G.~Narain, ``{On Gauss-bonnet gravity and boundary conditions in Lorentzian
  path-integral quantization},''
  \href{http://dx.doi.org/10.1007/JHEP05(2021)273}{{\em JHEP} {\bfseries 05}
  (2021) 273}, \href{http://arxiv.org/abs/2101.04644}{{\ttfamily
  arXiv:2101.04644 [gr-qc]}}.

\bibitem{Narain:2022msz}
G.~Narain, ``{Surprises in Lorentzian path-integral of Gauss-Bonnet gravity},''
  \href{http://dx.doi.org/10.1007/JHEP04(2022)153}{{\em JHEP} {\bfseries 04}
  (2022) 153}, \href{http://arxiv.org/abs/2203.05475}{{\ttfamily
  arXiv:2203.05475 [gr-qc]}}.

\bibitem{Ailiga:2023wzl}
M.~Ailiga, S.~Mallik, and G.~Narain, ``{Lorentzian Robin Universe},''
  \href{http://dx.doi.org/10.1007/JHEP01(2024)124}{{\em JHEP} {\bfseries 01}
  (2024) 124}, \href{http://arxiv.org/abs/2308.01310}{{\ttfamily
  arXiv:2308.01310 [gr-qc]}}.

\bibitem{Ailiga:2024mmt}
M.~Ailiga, S.~Mallik, and G.~Narain, ``{Lorentzian path-integral of Robin
  Universe},'' \href{http://arxiv.org/abs/2407.16692}{{\ttfamily
  arXiv:2407.16692 [gr-qc]}}.

\bibitem{Feldbrugge:2017fcc}
J.~Feldbrugge, J.-L. Lehners, and N.~Turok, ``{No smooth beginning for
  spacetime},'' \href{http://dx.doi.org/10.1103/PhysRevLett.119.171301}{{\em
  Phys. Rev. Lett.} {\bfseries 119} no.~17, (2017) 171301},
  \href{http://arxiv.org/abs/1705.00192}{{\ttfamily arXiv:1705.00192
  [hep-th]}}.

\bibitem{Feldbrugge:2018gin}
J.~Feldbrugge, J.-L. Lehners, and N.~Turok, ``{Inconsistencies of the New
  No-Boundary Proposal},''
  \href{http://dx.doi.org/10.3390/universe4100100}{{\em Universe} {\bfseries 4}
  no.~10, (2018) 100}, \href{http://arxiv.org/abs/1805.01609}{{\ttfamily
  arXiv:1805.01609 [hep-th]}}.

\bibitem{DiazDorronsoro:2018wro}
J.~Diaz~Dorronsoro, J.~J. Halliwell, J.~B. Hartle, T.~Hertog, O.~Janssen, and
  Y.~Vreys, ``{Damped perturbations in the no-boundary state},''
  \href{http://dx.doi.org/10.1103/PhysRevLett.121.081302}{{\em Phys. Rev.
  Lett.} {\bfseries 121} no.~8, (2018) 081302},
  \href{http://arxiv.org/abs/1804.01102}{{\ttfamily arXiv:1804.01102 [gr-qc]}}.

\bibitem{Halliwell:2018ejl}
J.~J. Halliwell, J.~B. Hartle, and T.~Hertog, ``{What is the No-Boundary Wave
  Function of the Universe?},''
  \href{http://dx.doi.org/10.1103/PhysRevD.99.043526}{{\em Phys. Rev. D}
  {\bfseries 99} no.~4, (2019) 043526},
  \href{http://arxiv.org/abs/1812.01760}{{\ttfamily arXiv:1812.01760
  [hep-th]}}.

\bibitem{Janssen:2019sex}
O.~Janssen, J.~J. Halliwell, and T.~Hertog, ``{No-boundary proposal in biaxial
  Bianchi IX minisuperspace},''
  \href{http://dx.doi.org/10.1103/PhysRevD.99.123531}{{\em Phys. Rev. D}
  {\bfseries 99} no.~12, (2019) 123531},
  \href{http://arxiv.org/abs/1904.11602}{{\ttfamily arXiv:1904.11602 [gr-qc]}}.

\bibitem{Vilenkin:2018dch}
A.~Vilenkin and M.~Yamada, ``{Tunneling wave function of the universe},''
  \href{http://dx.doi.org/10.1103/PhysRevD.98.066003}{{\em Phys. Rev. D}
  {\bfseries 98} no.~6, (2018) 066003},
  \href{http://arxiv.org/abs/1808.02032}{{\ttfamily arXiv:1808.02032 [gr-qc]}}.

\bibitem{Vilenkin:2018oja}
A.~Vilenkin and M.~Yamada, ``{Tunneling wave function of the universe II: the
  backreaction problem},''
  \href{http://dx.doi.org/10.1103/PhysRevD.99.066010}{{\em Phys. Rev. D}
  {\bfseries 99} no.~6, (2019) 066010},
  \href{http://arxiv.org/abs/1812.08084}{{\ttfamily arXiv:1812.08084 [gr-qc]}}.

\bibitem{Bojowald:2018gdt}
M.~Bojowald and S.~Brahma, ``{Loops rescue the no-boundary proposal},''
  \href{http://dx.doi.org/10.1103/PhysRevLett.121.201301}{{\em Phys. Rev.
  Lett.} {\bfseries 121} no.~20, (2018) 201301},
  \href{http://arxiv.org/abs/1810.09871}{{\ttfamily arXiv:1810.09871 [gr-qc]}}.

\bibitem{DiTucci:2018fdg}
A.~Di~Tucci and J.-L. Lehners, ``{Unstable no-boundary fluctuations from sums
  over regular metrics},''
  \href{http://dx.doi.org/10.1103/PhysRevD.98.103506}{{\em Phys. Rev. D}
  {\bfseries 98} no.~10, (2018) 103506},
  \href{http://arxiv.org/abs/1806.07134}{{\ttfamily arXiv:1806.07134 [gr-qc]}}.

\bibitem{Lehners:2021jmv}
J.-L. Lehners, ``{Wave function of simple universes analytically continued from
  negative to positive potentials},''
  \href{http://dx.doi.org/10.1103/PhysRevD.104.063527}{{\em Phys. Rev. D}
  {\bfseries 104} no.~6, (2021) 063527},
  \href{http://arxiv.org/abs/2105.12075}{{\ttfamily arXiv:2105.12075
  [hep-th]}}.

\bibitem{Matsui:2022lfj}
H.~Matsui, S.~Mukohyama, and A.~Naruko, ``{No smooth spacetime in Lorentzian
  quantum cosmology and trans-Planckian physics},''
  \href{http://dx.doi.org/10.1103/PhysRevD.107.043511}{{\em Phys. Rev. D}
  {\bfseries 107} no.~4, (2023) 043511},
  \href{http://arxiv.org/abs/2211.05306}{{\ttfamily arXiv:2211.05306 [gr-qc]}}.

\bibitem{Matsui:2024bfn}
H.~Matsui, ``{No smooth spacetime: Exploring primordial perturbations in
  Lorentzian quantum cosmology},''
  \href{http://dx.doi.org/10.1103/PhysRevD.110.023503}{{\em Phys. Rev. D}
  {\bfseries 110} no.~2, (2024) 023503},
  \href{http://arxiv.org/abs/2404.18609}{{\ttfamily arXiv:2404.18609 [gr-qc]}}.

\bibitem{Ailiga:2024nkz}
M.~Ailiga, S.~Mallik, and G.~Narain, ``{Boundary choices and one-loop Complex
  Universe},'' \href{http://arxiv.org/abs/2410.19724}{{\ttfamily
  arXiv:2410.19724 [gr-qc]}}.

\end{thebibliography}\endgroup

\end{document}